\documentclass[%
 reprint,
 superscriptaddress,
 showpacs,
 showkeys,
 amsmath,amssymb,
 aps,
 floatfix,
]{revtex4-1}

\usepackage{graphicx}
\usepackage{float}
\usepackage[utf8]{inputenc}
\usepackage{bm}
\usepackage{hyperref}
\usepackage{tabularx}
\usepackage{xcolor}
\usepackage[capitalize]{cleveref}

\makeatletter
\def\@fnsymbol#1{\ensuremath{\ifcase#1\or \dagger\or \ddagger\or
   \mathsection\or \mathparagraph\or \|\or **\or \dagger\dagger
   \or \ddagger\ddagger \else\@ctrerr\fi}}
    \makeatother
\newcommand{\deceased}[1]{\altaffiliation{#1}}

\graphicspath{{./graphics/}}

\begin{document}

\title{Equation of state for hot QCD and compact stars from a mean field approach}

\author{Anton Motornenko}
\affiliation{
Institut f\"ur Theoretische Physik,
Goethe Universit\"at, D-60438 Frankfurt am Main, Germany}
\affiliation{
Frankfurt Institute for Advanced Studies, Giersch Science Center,
D-60438 Frankfurt am Main, Germany}
\author{Jan Steinheimer}
\affiliation{
Frankfurt Institute for Advanced Studies, Giersch Science Center,
D-60438 Frankfurt am Main, Germany}
\author{Volodymyr Vovchenko}
\affiliation{
Institut f\"ur Theoretische Physik,
Goethe Universit\"at, D-60438 Frankfurt am Main, Germany}
\affiliation{
Frankfurt Institute for Advanced Studies, Giersch Science Center,
D-60438 Frankfurt am Main, Germany}
\author{Stefan Schramm}
\deceased{Deceased.}
\affiliation{
Institut f\"ur Theoretische Physik,
Goethe Universit\"at, D-60438 Frankfurt am Main, Germany}
\affiliation{
Frankfurt Institute for Advanced Studies, Giersch Science Center,
D-60438 Frankfurt am Main, Germany}
\author{Horst Stoecker}
\affiliation{
Institut f\"ur Theoretische Physik,
Goethe Universit\"at, D-60438 Frankfurt am Main, Germany}
\affiliation{
Frankfurt Institute for Advanced Studies, Giersch Science Center,
D-60438 Frankfurt am Main, Germany}
\affiliation{
GSI Helmholtzzentrum f\"ur Schwerionenforschung GmbH, D-64291 Darmstadt, Germany}

\date{\today}

\begin{abstract}
The thermodynamic properties of high temperature and high density QCD-matter
are explored within the Chiral SU(3)-flavor parity-doublet Polyakov-loop quark-hadron mean-field model, CMF. 
The quark sector of the CMF model is tuned to describe the $\mu_B=0$ thermodynamics data of lattice QCD. 
The resulting lines of constant physical variables as well as the baryon number susceptibilities  are studied in some detail in the temperature/chemical potential plane. 
The CMF model predicts three consecutive transitions, the nuclear first-order liquid-vapor phase transition, chiral symmetry restoration, and the cross-over transition to a quark-dominated phase. All three phenomena are cross-over, for most of the $T-\mu_B$-plane.
The deviations from the free ideal hadron gas baseline at $\mu_B=0$ and $T\approx 100-200$ MeV can be attributed to remnants of the liquid-vapor first order phase transition in nuclear matter. The chiral crossing transition determines the baryon fluctuations at much higher $\mu_B\approx1.5$ GeV, and at even higher baryon densities $\mu_B\approx2.4$ GeV, the behavior of fluctuations is controlled by the deconfinement cross-over.
The CMF model also describe well the static properties of high $\mu_B$ neutron stars as well as the new neutron star merger observations. 
The effective EoS presented here describes simultaneously lattice QCD results at $\mu_B=0$, as well as observed physical phenomena (nuclear matter and neutron star matter) at $T\cong0$ and high densities, $\mu_B>1$ GeV.
\end{abstract}

\maketitle

\section{\label{sec:int}Introduction}
At low temperatures and high baryonic densities a transition from hadronic matter to a cold deconfined state with quark degrees of freedom is expected~\cite{McLerran:2018hbz}, that is supported by perturbative QCD calculations~\cite{Kurkela:2009gj}. This transition is a particular case of appearance of a deconfined state of quark-gluon plasma (QGP) that expected at large densities and large temperatures~\cite{Shuryak:1980tp}. 
The study of the unknown QCD phase diagram is one of the main motivations for the state-of-the-art research in the nuclear and particle physics community.
A rich phase structure is conjectured for finite temperatures and chemical potentials~\cite{Rajagopal:1999cp, Alford:2002ng, Buballa:2003qv, Schafer:2003vz, Fukushima:2010bq}.
The QCD phase structure is a most important ingredient for the understanding of the Early Universe, ultra relativistic heavy ion collisions, 
and of the evolution, structure and inspirals of neutron stars~\cite{Takami:2015gxa}. 
Even though QCD is a well-established theory with only a few parameters, perturbative calculations are inappropriate in the crossover regions of $T$ and $\mu_B$ discussed here due to the large values of the QCD coupling constant at scales relevant for most of these applications~\cite{Braaten:1995jr}. The infamous sign problem \cite{1005.0539} prohibits lattice QCD (LQCD) calculations at finite densities.

The running experiments at the Large Hadron Collider (LHC), at the Relativistic Heavy Ion Collider (RHIC), at the Super Proton Synchrotron (SPS), and at the Heavy Ion Synchrotron (SIS) provide state-of-the-art data of the measurements of properties of matter produced in heavy-ion collisions (HIC). 
The detailed information of the particle production as measured in these experiments
allow to extract both thermodynamic and kinetic characteristics of the system that is created. 

Astrophysical observations of compact stars, along with data from the recent  gravitational-wave detection by LIGO provide an additional tool to probe the equation of state of dense nuclear and possible quark matter~\cite{1711.02644, 1803.00549, 1805.11581, Most:2018eaw, Bauswein:2018bma, Hanauske:2017oxo, Hanauske:2019qgs} in the region of moderate temperatures and high baryon densities, close to those as created in HIC. 

First principle LQCD calculations suggest a smooth crossover transition at vanishing baryochemical potential $\mu_{\rm B}=0$ from hadronic to partonic degrees of freedom~\cite{Aoki:2006we}. Although there is no indication of a first or a second order phase transition in the energy density, pressure, entropy density and speed of sound, there are other observable in LQCD -- chiral susceptibilities which seem to indicate a chiral cross-over at a pseudocritical temperature $T_{\rm pc}\approx 155$ MeV \cite{Borsanyi:2013bia,Bazavov:2014pvz}. However, the extension of LQCD calculations to finite $\mu_{\rm B}$ is a difficult problem. 
There are approximate lattice methods, such as extrapolations by Taylor expansion and the analytic continuation from purely imaginary to real $\mu_B$ are reasonable only for small baryon densities.
The exploration of higher baryon densities require effective QCD models, which respect the known symmetries of QCD and describe appropriately the known phenomenology of strong interactions.

The current knowledge of the properties of strongly interacting matter suggests a number of features that ought to be incorporated in any reasonable effective QCD model:
\begin{itemize}
    \item First, nuclear matter and the nuclear liquid-vapor phase transition at moderate temperatures, close to the nuclear saturation density~\cite{Pochodzalla:1995xy};
    \item Second, the chiral symmetry restoration should lead to the Stefan-Boltzmann limit for the thermodynamic properties at high temperature and at high chemical potential~\cite{Borsanyi:2010bp};
    \item The transition from hadronic- to quark-gluon degrees of freedom, at high temperatures and/or chemical potentials, is a crucial ingredient for the consistent description of QCD matter.
\end{itemize}
Often, these different aspects of QCD are modeled within separate frameworks, which are then merged through various constructions.

The present work formulates a single combined framework to describing the QCD thermodynamics, which simultaneously satisfies all the constraints, from lattice QCD, and known nuclear matter properties, as well as neutron star observations. 

The resulting equation of state is then used to estimate various properties of systems created in both heavy-ion collisions and neutron star physics.
Sec. \ref{sec:model} presents the description of the Chiral SU(3)-flavor parity-doublet Polyakov-loop quark-hadron mean field model, CMF.
Section \ref{sec:fit} describes the fine-tuning of model parameters to the $\mu_{\rm B}=0$ LQCD data on the trace anomaly, and presents a comparison with the CMF model predictions for various conserved charge number fluctuations with the corresponding lattice data.
The QCD phase diagram deduced from the CMF model is studied in Sec.~\ref{sec:ph-dia}.
The creation of hot and dense QCD matter as created in heavy-ion collisions at various collision energies is studied using the 1-dimensional hydrodynamics in Sec.~\ref{sec:HI-coll}, the respective trajectories along the QCD phase diagram are explored as well. 
Section \ref{sec:NS} presents the CMF model predictions for the observed neutron star properties.

\section{\label{sec:model}Chiral SU(3)-flavor parity-doublet Polyakov-loop quark-hadron mean-field model, CMF}

The Chiral SU(3)-flavor parity-doublet Polyakov-loop quark-hadron mean-field model, CMF, is an extension of the previously proposed $\sigma$-$\omega$ model with parity doubling for nuclear and hadron matter~\cite{Detar:1988kn,Hatsuda:1988mv,Papazoglou:1996hf,Papazoglou:1997uw, Papazoglou:1998vr,Sasaki:2010bp}. The CMF model was extended to include quark degrees of freedom~\cite{Steinheimer:2010ib, Steinheimer:2011ea,Dexheimer:2012eu,Mukherjee:2016nhb, Motornenko:2018hjw}. This model is a phenomenological effective unified approach to describe interacting hadron-quark matter. The Lagrangian includes essential symmetries and features of QCD. These include:
\begin{itemize}
\item Chiral symmetry restoration in the hadronic sector, in particular the baryon parity doubling 
so an explicit mass term for baryons is possible, even when chiral symmetry is restored. 
This leads to a restoration of mass degeneracy among baryons and their respective parity partners~\cite{Aarts:2017rrl,Aarts:2018glk}.
\item Eigenvolume corrections for hadrons, which allow for an effective modeling of their repulsive interactions.
This suppresses hadronic densities and ensures a transition to a parton-dominated matter at large densities, when quark and gluon d.o.f. appear.
\item Chiral symmetry restoration for quarks and a dynamical generation of their masses.
\item The Polyakov loop via a QCD-motivated potential incorporates the deconfinement transition.
\end{itemize}

A detailed description of the hadronic part of the CMF model can be found
in the literature \cite{Dexheimer:2007tn,Steinheimer:2010ib, Steinheimer:2011ea, Mukherjee:2016nhb}. 
It is based on a realization of a $\sigma$-$\omega$ model in mean-field description. Here the relevant fermionic degrees of freedom are baryons that interact through mesonic mean-fields. 
The version of this model used here includes all states in the SU(3)$_f$ baryon octet, together with their parity partners, i.e. states with the same quantum numbers but opposite parity. The LQCD data suggests that the same mechanism should be implemented for higher baryonic states -- the baryon octet \cite{Aarts:2018glk}, this is a plan for future studies. In the limit of chiral symmetry restoration these parity partner states should be degenerate and their masses equal, which then serves as a signal for chiral symmetry restoration. 
To allow for such a behavior, the baryon masses are dynamically generated by their couplings to the scalar $\sigma$-field and the scalar strange $\zeta$-field, which serve as the order parameters for the chiral transition:
\begin{eqnarray}
m^*_{\text{i}\pm} &=& \sqrt{ \left[ (g^{(1)}_{\sigma \text{i}} \sigma + g^{(1)}_{\zeta \text{i}}  \zeta )^2 + (m_0+n_\text{s} m_\text{s})^2 \right]}\nonumber \\
& \pm & g^{(2)}_{\sigma \text{i}} \sigma \pm g^{(2)}_{\zeta \text{i}} \zeta ~.
\label{effmass}
\end{eqnarray}
Here $+$ stand for positive and $-$ for negative parity states, $g^\text{(j)}_\text{i}$ are the coupling constants of baryons to the two scalar fields, $m_0=759$ MeV is the baryon mass at the restored phase. In addition, there is an SU(3)$_f$ symmetry-breaking mass term proportional to the strangeness content of the baryons, there $n_s$ is the number of strange quarks in the baryon, and $m_s=130$ MeV is the mass of the strange quark. The couplings $g^\text{(j)}_\text{i}$ are tuned to reproduce the vacuum masses of baryons.

The mean-field values of the chiral fields are driven by the thermal contribution from baryons and quarks, and controlled by the scalar meson interaction, driving the spontaneous breaking of the chiral symmetry:
\begin{equation}
V = V_0 + \frac{1}{2} k_0 I_2 - k_1 I_2^2 - k_2 I_4 + k_6 I_6 ~,
\label{veff}
\end{equation}
with
\begin{eqnarray}
    I_2 = (\sigma^2+\zeta^2)&,&~ I_4 = -(\sigma^4/2+\zeta^4),\nonumber\\
    I_6 &=& (\sigma^6 + 4\, \zeta^6)
\end{eqnarray}
where $V_0$ is fixed by demanding that the potential vanishes in the vacuum.\\ 

The parameters of the scalar and the vector interactions are fitted to describe nuclear matter properties~\cite{Steinheimer:2014kka, Mukherjee:2016nhb}.
Contributions of all established hadronic resonances are included here with their vacuum masses~\cite{Tanabashi:2018oca}. 
These states can be coupled to meson fields as parity doublets as well. However, this is not done in the current implementation, they only interact with the other particles via their excluded volume only.\\ 

The quark degrees of freedom are incorporated similarly with PNJL approach~\cite{Fukushima:2003fw}. 
The appearance of quarks is controlled by the value of Polyakov loop $\Phi$, which plays the role of the order parameter for the deconfinement transition.
The coupling of the quarks to the Polyakov loop is introduced through the thermal energy of the quarks.
Their thermal contribution to the grand canonical potential $\Omega$ is given by
\begin{equation}
	\Omega_{q}=-T \sum_{i\in Q}{\frac{d_i}{(2 \pi)^3}\int{d^3k \ln\left(1+\Phi \exp{\frac{-\left(E_i^*-\mu^*_i\right)}{T}}\right)}}
\end{equation}
and
\begin{equation}
	\Omega_{\overline{q}}=-T \sum_{i\in Q}{\frac{d_i}{(2 \pi)^3}\int{d^3k \ln\left(1+\Phi^* \exp{\frac{-\left(E_i^*+\mu^*_i\right)}{T}}\right)}}.
\end{equation}
The sums run over all light quark flavors (u, d, and s), $d_i$ is the corresponding degeneracy factor, $E^*_i=\sqrt{m^{*2}_i+p^2}$ is the energy, and $\mu^*$ is the chemical potential of the quark. Note, two- and three- quark contributions to $\Omega$ are omitted in the CMF model since hadronic excitations are explicitly included.

The effective masses of the light quarks are generated by the $\sigma$ field (non-strange chiral condensate) as well, the mass of the strange quark is generated by the $\zeta$ field (strange quark-antiquark state). The small explicit mass terms $\delta m_q=5$ MeV, and $\delta m_s=150$ MeV for the strange quark, and $m_0q=253$ MeV which corresponds to an explicit mass term which does not originate from chiral symmetry breaking:
\begin{eqnarray}
m_{q}^* & =-g_{q\sigma}\sigma+\delta m_q + m_{0q},&\nonumber\\
m_{s}^* & =-g_{s\zeta}\zeta+\delta m_s + m_{0q}.&
\end{eqnarray}

Dynamics of the Polyakov-loop is controlled by the effective Polyakov-loop potential $U(\Phi, \Phi^*, T)$~\cite{Ratti:2005jh}:
\begin{align}
	U&=-\frac12 a(T)\Phi\Phi^*\nonumber\\
	&\quad + b(T)\log[1-6\Phi\Phi^*+4(\Phi^3 + \Phi^{*3})-3(\Phi\Phi^*)^2],\nonumber\\
    a(T)&=a_0 T^4+a_1 T_0 T^3+a_2 T_0^2 T^2,~
    b(T)=b_3 T_0^4
\end{align}

The parameters of this potential can be fixed to the lattice QCD data in the pure gauge sector~\cite{Ratti:2005jh}. However, this yields an unsatisfactory description of the (2+1)-flavor QCD thermodynamics when the hadrons are explicitly included in the model. Therefore the parameters of the Polyakov Loop potential are adjusted in the present work to describe properly the (2+1)-flavor lattice data.\\

The CMF model incorporates excluded-volume effects, in order to suppress the hadronic degrees of freedom in the regions of the phase diagram where physically quarks and gluons dominate~\cite{Steinheimer:2011ea}.
Consequently, all the thermodynamics densities, including the quark contribution, are reduced as parts of the system are occupied by EV-hadrons:
\begin{eqnarray}
\rho_i=\frac{\rho^{\rm id}_i (T, \mu^*_i - v_i\,p)}{1+\sum\limits_j v_j \rho^{\rm id}_j (T, \mu^*_j - v_j\,p)},
\end{eqnarray}
the $v_j$ are the eigenvolume parameters for the different species. $p$ is a system pressure without contribution of mean fields, $\mu^*$ is the chemical potential of the hadron.
The $v$ is assumed to be $v_B = 1$~fm$^3$ for (anti-) baryons, $v_M = 1/8$~fm$^3$ for mesons, and is set to zero $v_q = 0$ for quarks.

\section{\label{sec:fit}Constraining the CMF model to the lattice data}
\begin{figure}[H]
\centering
\includegraphics[width=.49\textwidth]{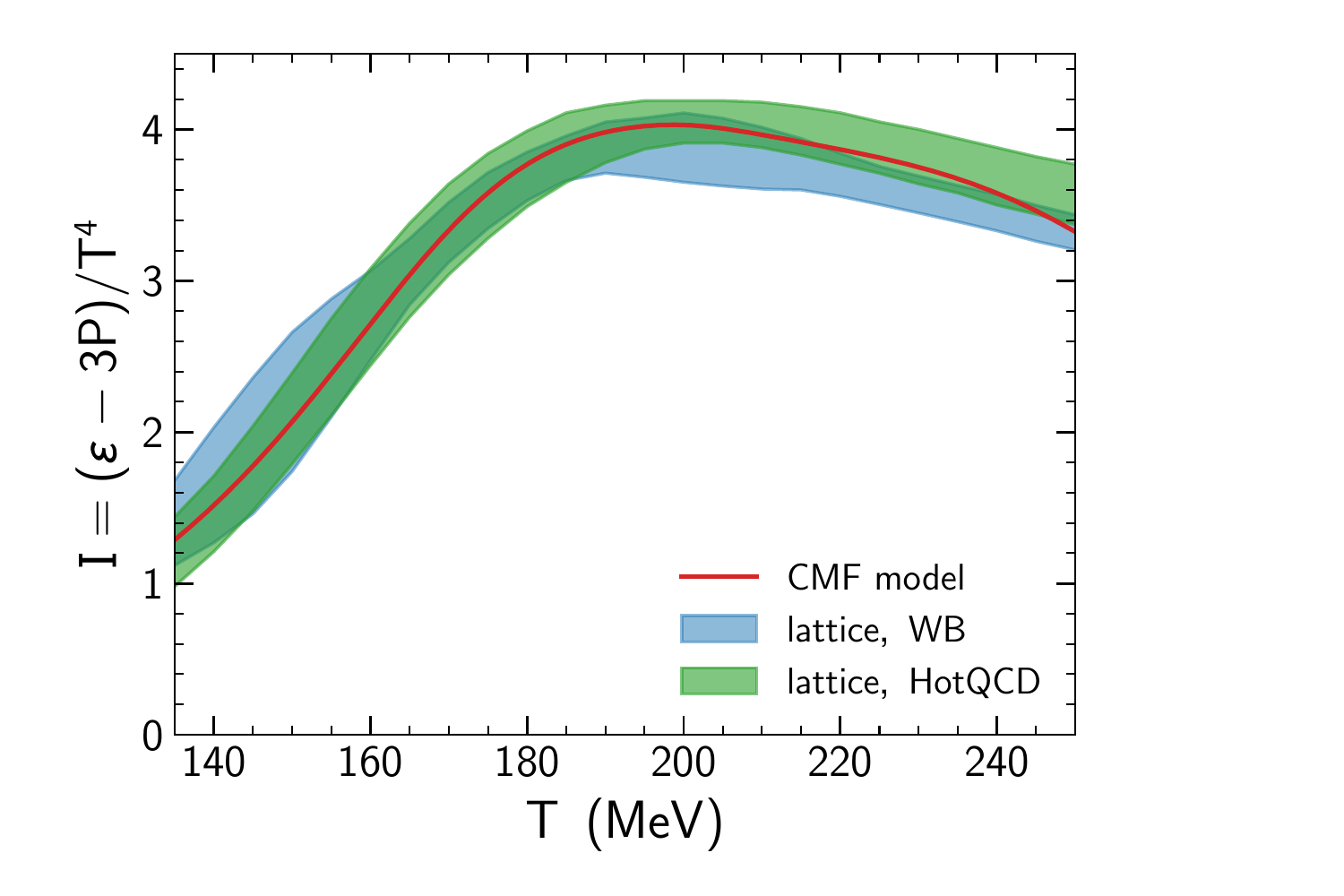}
\caption{Trace anomaly $I$ at $\mu_B=0$ as a function of temperature $T$. Comparison between model predictions and LQCD results \cite{Borsanyi:2013bia, Bazavov:2014pvz}.}
\label{fig:fit}
\end{figure}
To introduce the constraints on the CMF model parameters from lattice QCD at high temperature and zero net-baryon density we use the QCD trace anomaly $I$, ``the interaction measure'', as a reference:
\begin{eqnarray}
\frac{I}{T^4}=\frac{\varepsilon - 3P}{T^4}\, .
\end{eqnarray}
The free parameters of the present model are tuned to reproduce this quantity.
The LQCD trace anomaly permits to calculate all observable thermodynamic quantities. Analysis of other Lattice data (chiral susceptibility) seem to show that for chirally related observables there is a crossover transition with a ``pseudo''-critical temperature at $T\approx156$ MeV. The analysis of this data by a phenomenological model suggests a half-hadron, half-quark composition in that region \cite{Albright:2014gva}.

The parameters of the CMF model's quark sector needed to reproduce the trace anomaly data from LQCD are found by a least mean squares fitting procedure for the parameters of the Polyakov loop potential $U(\Phi, \Phi^*, T)$ and for the coupling constants $g_{q\sigma}$ and $g_{s\zeta}$ of the quarks to the chiral condensates $\sigma$ and $\zeta$, respectively. 
All in all this fixes 5 model parameters, $T_0, a_1, a_2$, $b_3,~g_{q\sigma}=g_{s\zeta}$ (we set $g_{q\sigma}$ and $g_{s\zeta}$ to the same value).
The quark parameter fitting is performed through a scan over the parameter space on a $8\times6\times7\times6\times6$ sized grid, minimizing the root mean square deviation of the CMF model data on $I/T^4$ and computed on the lattice results. The resulting parameter values are presented in Table~\ref{tabl:params}.
The comparison of the CMF model to the lattice data is shown in Fig.~\ref{fig:fit}.
\begin{table}[H]
\centering
\begin{tabular*}{0.45\textwidth}{@{\extracolsep{\fill}} ccccc}
$T_0$ (MeV) & $a_1$ & $a_2$ & $b_3$ & $g_{q\sigma}=g_{s\zeta}$\\
\hline
\hline
180.0 & -11.67 & 9.33 & -0.53 & -1.0
\end{tabular*}
\caption{Best fit values of parameters extracted from a scan over the parameter space.}
\label{tabl:params}
\end{table}

\begin{figure}[t]
\centering
\includegraphics[width=.49\textwidth]{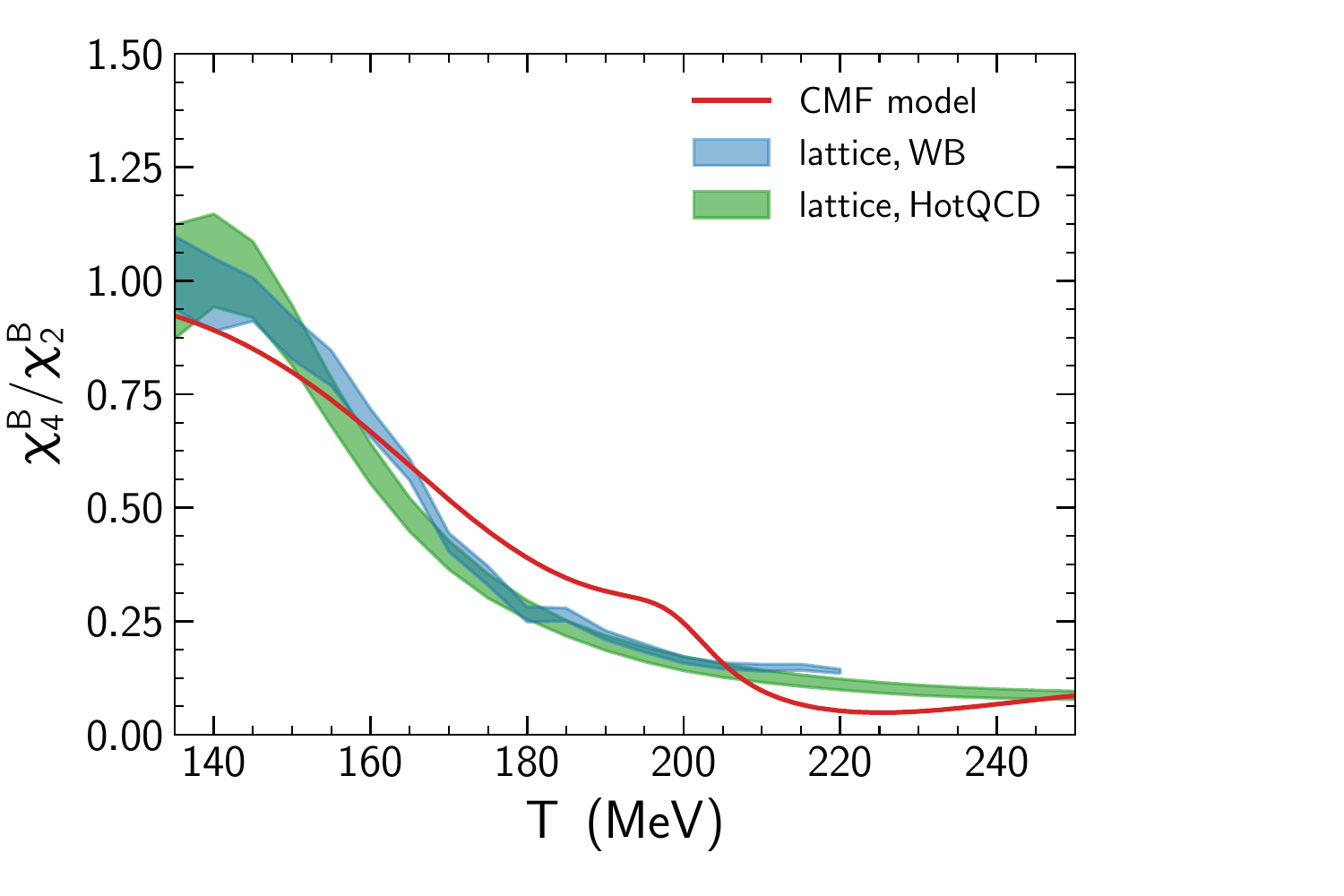}

\caption{The temperature dependence of net baryon number susceptibility ratio $\chi_4^B/\chi_2^B$ is shown as a function of temperature $T$. The red line depicts calculations within the CMF model, blue and green color bands are the results of the  lattice QCD calculations from the Wuppertal-Budapest and the HotQCD collaborations, respectively \cite{Borsanyi:2011sw, Bellwied:2015lba, Bazavov:2017dus, Bazavov:2017tot}.}
\label{fig:mu0}
\end{figure}

The values in Table~\ref{tabl:params} suggest couplings $g_{q\sigma},~g_{s\zeta}$ of quarks to the chiral fields of about $1/3$ of the baryons, this one may expect from the additive quark model. Larger values of quark couplings would significantly influence the size of the peak in interaction measure $I/T^4$, as studied in Ref.~\cite{Mukherjee:2016nhb}. In case of large values of $g_{q\sigma}, g_{s\zeta}$ the strong interplay between the chiral symmetry restoration and the deconfinement transition would result in too large values of the interaction measure. The large peaks in the baryon number susceptibilities are in contrast to the lattice data.

Higher order baryon number susceptibilities $\chi_n^B$ that are a LQCD measure of the particle number fluctuations
\begin{eqnarray}
\chi_n^B= \frac{\partial^n (P/T^4)}{(\partial \mu_B/T)^n}\,,
\label{eq:susc}
\end{eqnarray}
as well as the curvatures of various lines of constant physical quantities are also interesting in scope of LQCD data.

The behavior of the $\chi_4^B/\chi_2^B$ at $\mu_B=0$ is presented in this section, the study for finite values of $\mu_B$ is presented in Sec. \ref{sec:ph-dia}.
The comparison of the CMF model with the available LQCD data for the $\chi_4^B/\chi_2^B$ is shown in Fig.~\ref{fig:mu0}, indicating a fair agreement of the CMF model with the lattice data.

Lattice QCD studies often explore regions of finite $\mu_B$ by using the Taylor series expansion. The Taylor expansion in series of $T$ and $\mu_B$ up to $\mathcal{O}(\mu_B^4)$ was used in \cite{Bazavov:2017dus} to calculate ``lines of constant physics'': i.e. lines in the $T-\mu_B$ plane where certain thermodynamic quantities like pressure, energy density, and entropy density $P, \varepsilon, s$, are constant. The coefficients $\kappa^f_2$ and $\kappa^f_4$ ($f\equiv P, \varepsilon, s$) represent these contour lines in the $T-\mu_B$ plane using the following parametrization~\cite{Bazavov:2017dus}:
\begin{equation}
\label{eq:LCP}
T_f(\mu_B) = T_0 \left(1-\kappa^f_2 
\left( \frac{\mu_B}{T_0}\right)^2- \kappa^f_4 \left( \frac{\mu_B}{T_0}
\right)^4\right)\,.
\end{equation}
Here the coefficients $\kappa^f_2$ and $\kappa^f_4$ are calculated from \cref{eq:k2,eq:k_help}, see Ref.~\cite{Bazavov:2017dus} for details:
\begin{widetext}
\begin{align}
\label{eq:k2}
\kappa^f_2 &=  \frac{1}{T_0}\frac{ f_2(T_0)}{
\left.  \frac{\partial f_0(T)}{\partial T}\right|_{(T_0, 0)} }, \qquad
\kappa^f_4 &= \frac{\frac12 (\kappa^f_2)^2 \left. T_0^2\frac{\partial^2 f_0(T)}{\partial T^2}\right|_{(T_0, 0)}  -
\kappa^f_2\left. 
\left( T_0 \frac{\partial f_2(T)}{\partial T} \right|_{(T_0, 0)} -2 f_2(T_0) \right)
  + f_4(T_0)}{\left. 
\frac{\partial f_0(T)}{\partial T}\right|_{(T_0, 0)} }.
\end{align}
\end{widetext}
Here 
\begin{equation}
\label{eq:k_help}
f_{2n}(T) = \frac{1}{(2n)!} \left. \frac{\partial^{2n}f(T, \mu_B)}{\partial \left(\mu_B/T\right)^{2n}} \right|_{(T,0)}\,.
\end{equation}

\begin{figure*}[t]
\centering
\includegraphics[width=.49\textwidth]{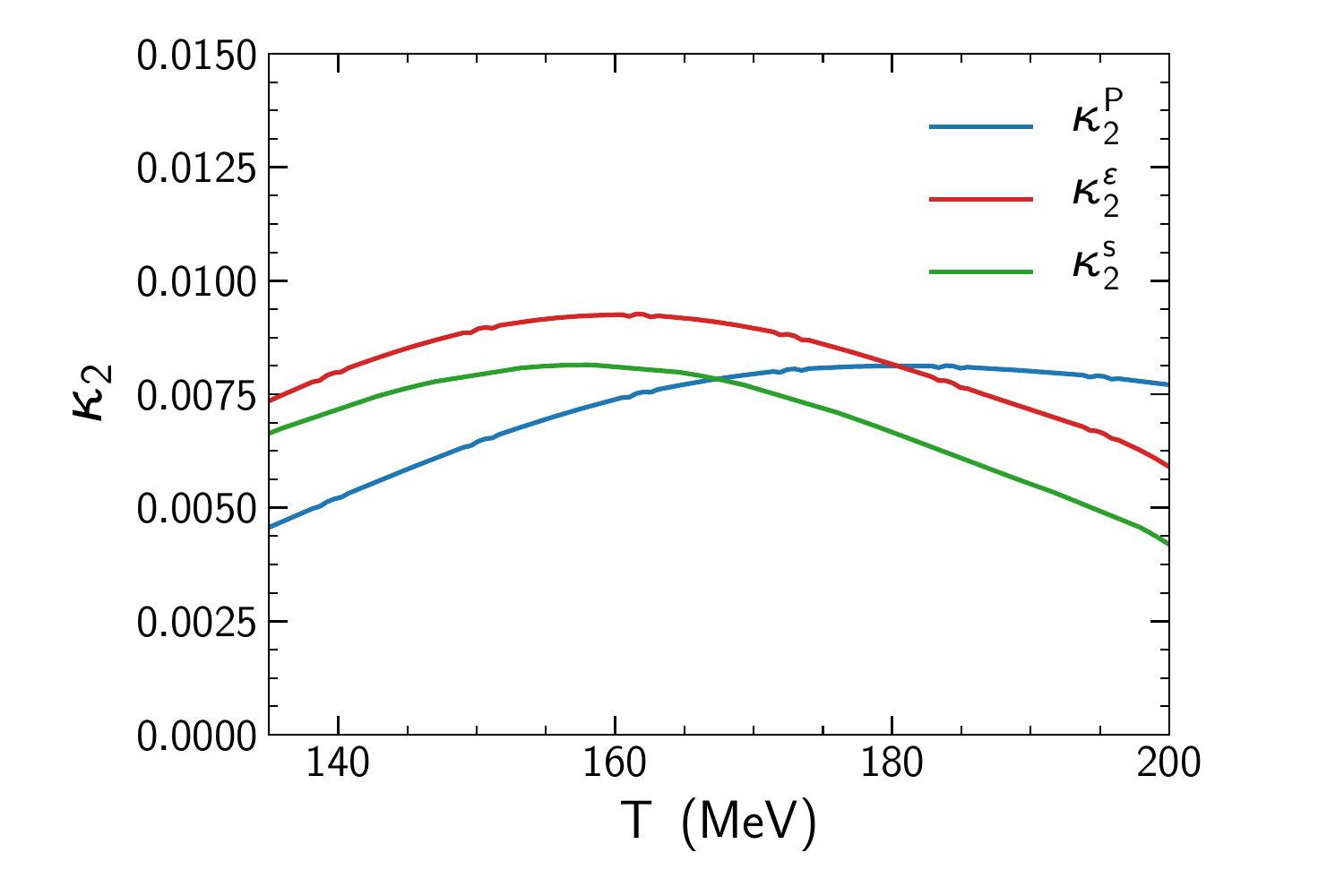}
\includegraphics[width=.49\textwidth]{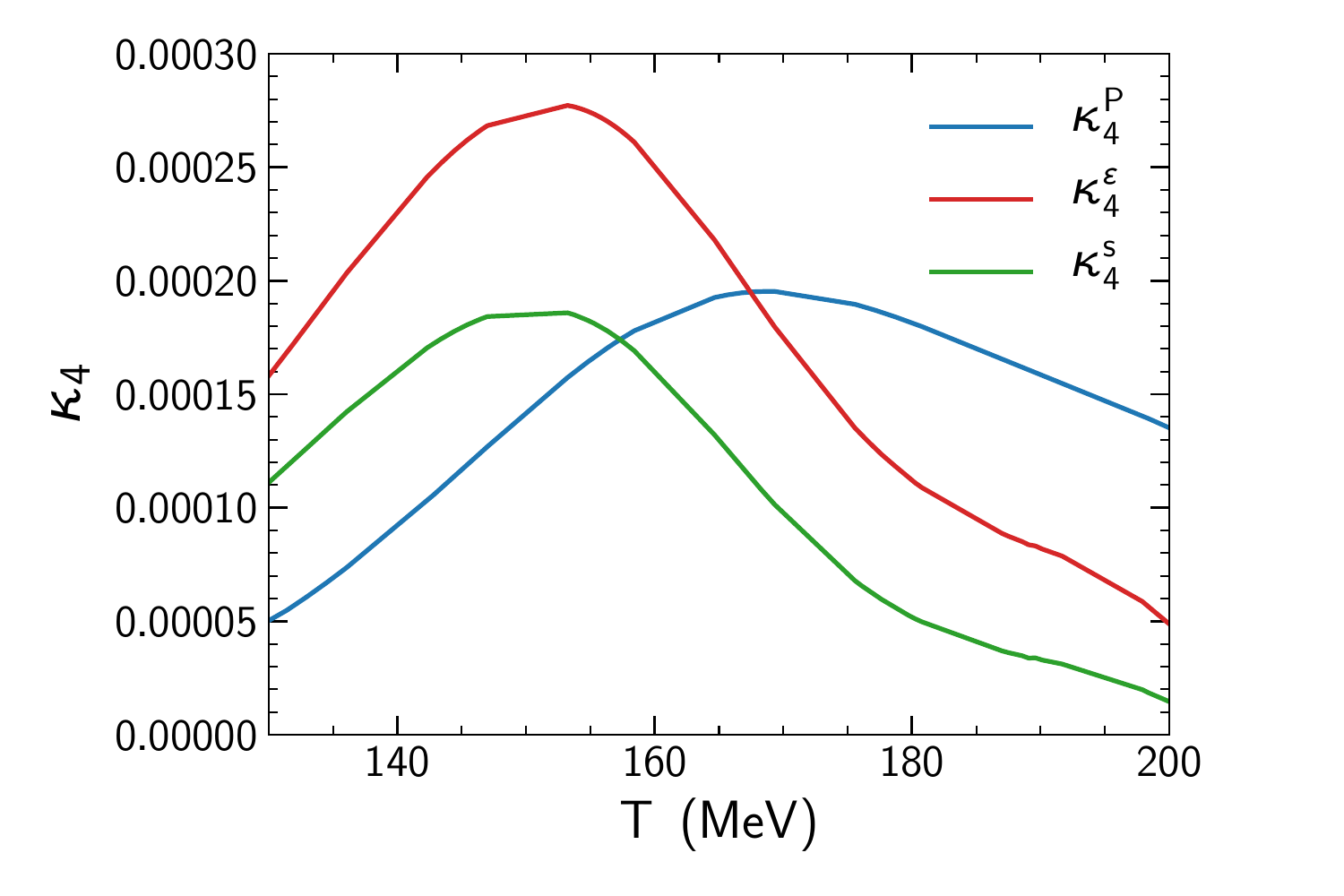}
\caption{Second $\kappa^f_2$ ({\bf left}) and fourth $\kappa^f_4$ ({\bf right}) coefficients of line of constant pressure $P$, energy density $\varepsilon$ and entropy density $s$ versus temperature $T$. Explanation and lattice data can be found at \cite{Bazavov:2017dus}.}
\label{fig:const-P}
\end{figure*}

The coefficients $\kappa^f_2$ and $\kappa^f_4$ are calculated in the CMF model for the pressure $P$, the energy density $\varepsilon$, and the entropy density $s$ as functions of the temperature $T$. The CMF model predictions are in a reasonable agreement with recent LQCD calculations~\cite{Bazavov:2017dus}. 
The rather low values of $\kappa_2^f$ and $\kappa_4^f$ suggest also small curvatures of lines of constant physical observables in the temperature region studied here. Effects of the finite chemical potential are small, therefore these lines are almost horizontal in the $T$-$\mu_B$ plane. 
The coefficients for the entropy and the energy density indicate that $\kappa_2^s < \kappa_2^\varepsilon$, meaning a decrease of the entropy density along the lines of constant energy density.

\section{\label{sec:ph-dia}The CMF model phase diagram}

\begin{figure}[t]
\centering
\includegraphics[width=.49\textwidth]{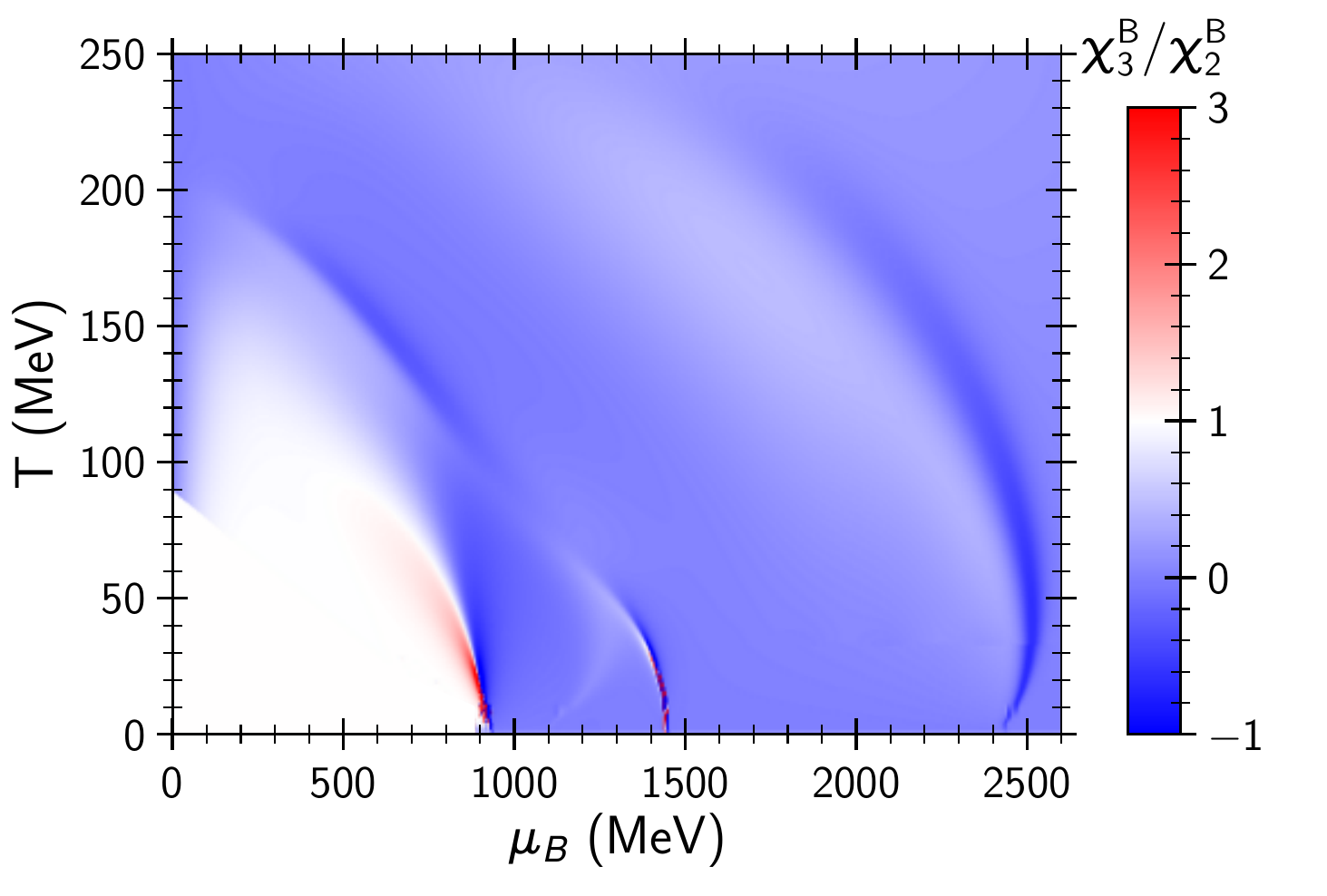}
\includegraphics[width=.49\textwidth]{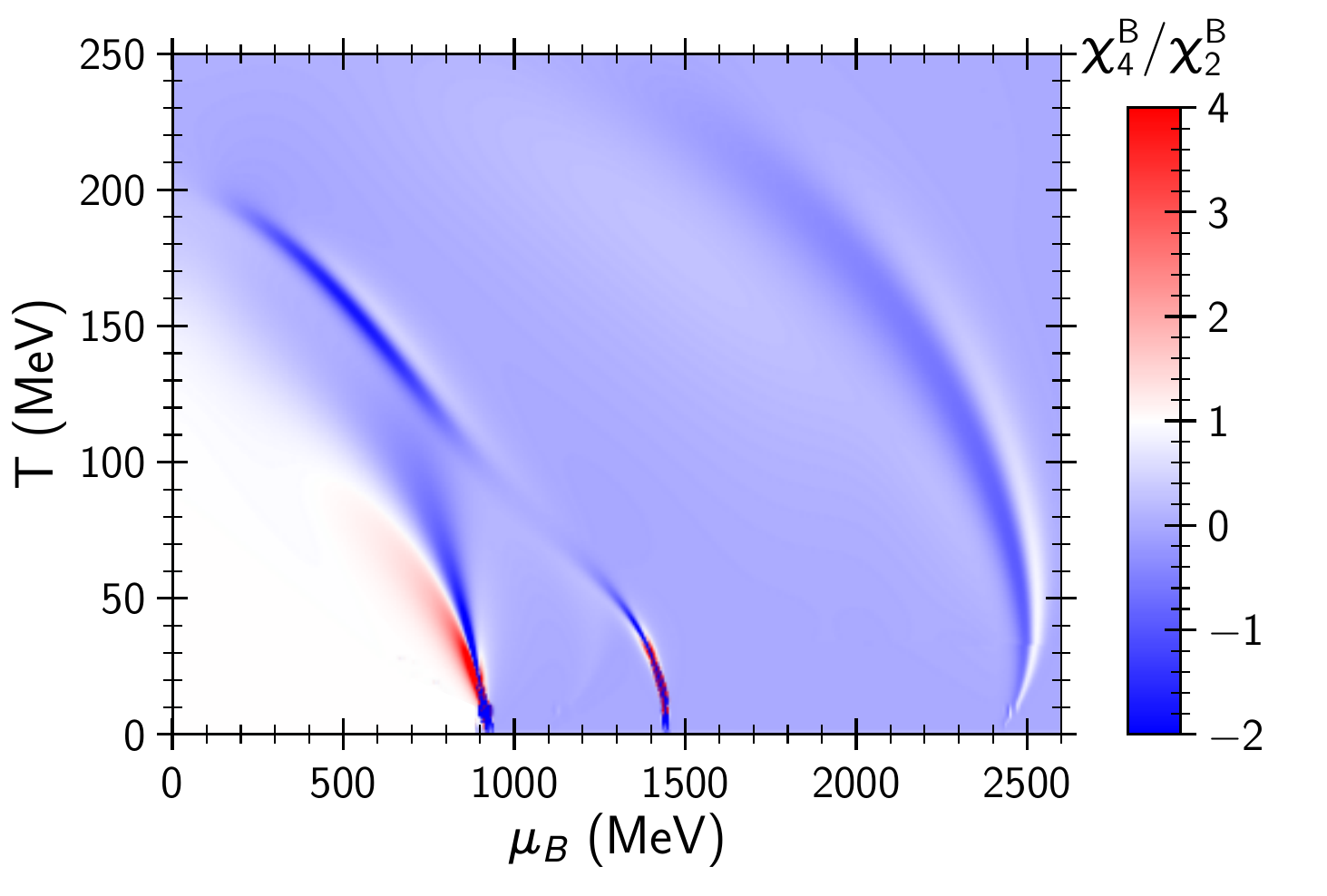}\\

\caption{Ratios of the CMF baryon number susceptibilities $\chi_3^{\rm B}/\chi_2^{\rm B}$ skewness ({\bf upper}) and $\chi_4^{\rm B}/\chi_2^{\rm B}$ kurtosis ({\bf lower}) in the baryon chemical potential - $\mu_{\rm B}$ and temperature - $T$ plane. Note the 3 distinct critical regions, with their remnants reaching from $T=0$ up to $T>200$ MeV.}
\label{fig:suscept-diag_muT}
\end{figure}

\begin{figure}[t]
\centering
\includegraphics[width=.49\textwidth]{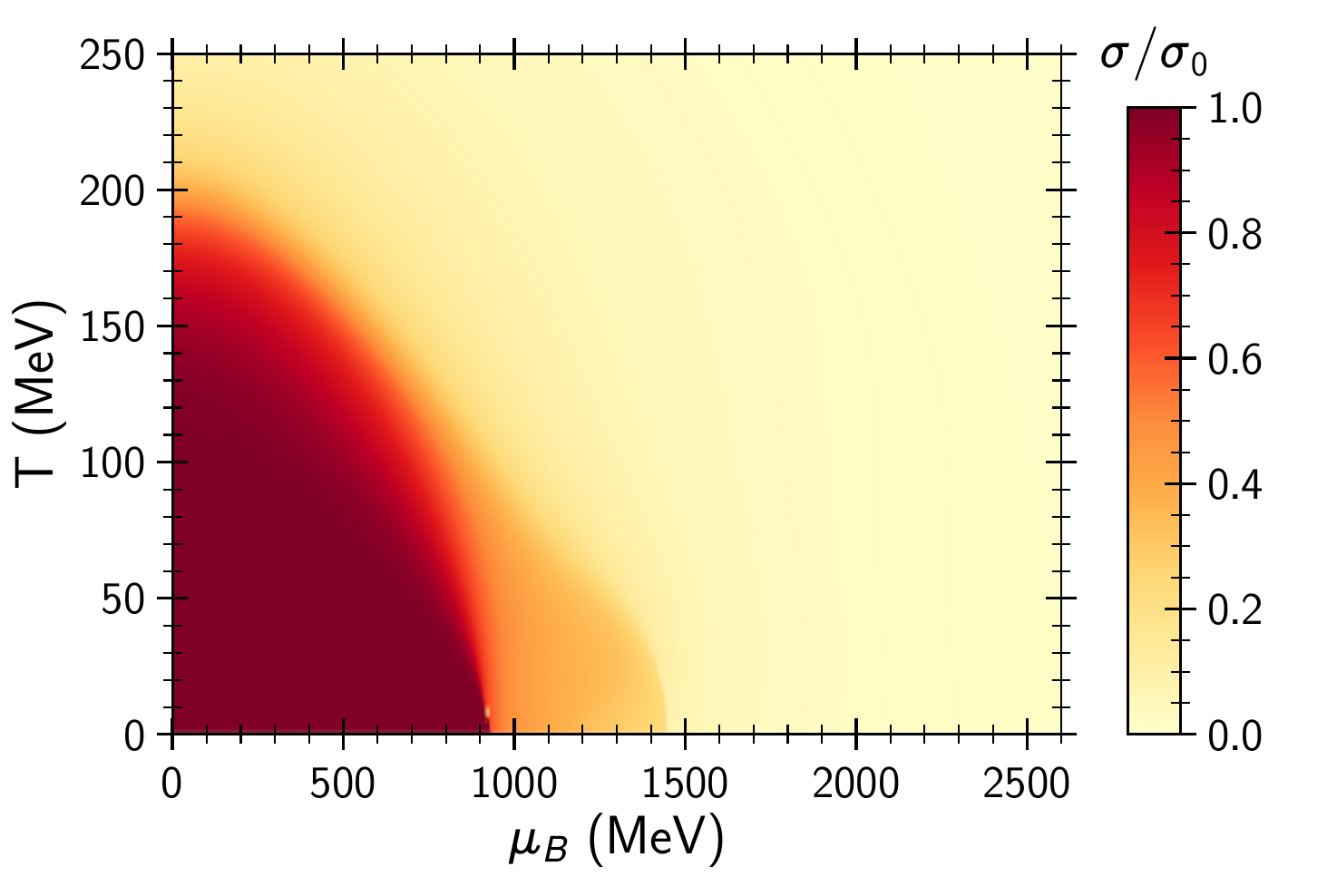}
\includegraphics[width=.49\textwidth]{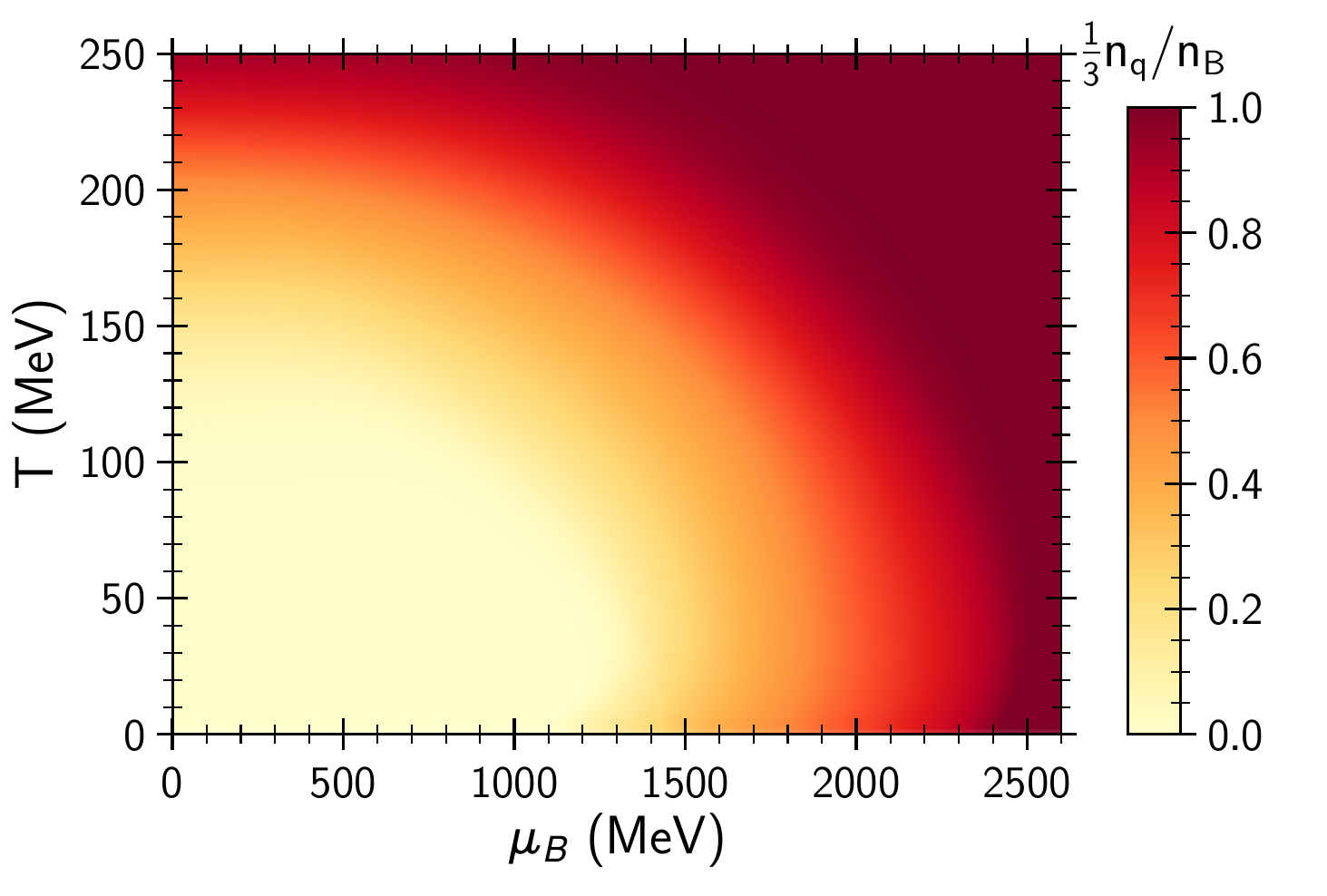}\\

  \caption{The normalized non-strange chiral condensate $\sigma/\sigma_0$ (the sigma field) ({\bf upper}) and the quark fraction $\frac13{n_q}/{n_B}$ ({\bf lower}) of the CMF model in baryon chemical potential $\mu_{\rm B}$ and temperature $T$ plane. Note that the rather fast change of the chiral condensate appears at moderate energy densities, while deconfinement appears only at much higher energy densities/chemical potentials.}
\label{fig:order-param_muT}
\end{figure}

Two order parameters, the chiral condensate $\sigma$ and the Polyakov loop $\Phi$, plus the interacting baryon octet within the SU(3)-flavor $\sigma$-model permit four different phases within the CMF model. 
These phases are characterized as:
\begin{itemize}
\item A dilute gas of interacting hadrons;
\item A hadronic liquid -- a dense hadronic phase, the transition from the hadron gas to the hadronic liquid is the nuclear liquid-vapor phase transition. Quarks start to appear in the hadronic liquid, but they are negligible;
\item A chirally restored phase, where the mass symmetry between the parity partners is restored. Here the quark masses are decreased, hence quarks give a sizable contribution to the thermodynamics;
\item A quark dominated phase, where most of the energy density is carried by quark and gluon degrees of freedom. The gluon contribution is modeled by the Polyakov loop potential~\cite{Ratti:2005jh} and $\epsilon_q/\epsilon_{\rm tot} \approx 1$.
\end{itemize}

The baryon number susceptibilities $\chi_n^B$, that can be calculated using Eq.~\ref{eq:susc}, are proportional to the respective cumulants of the baryon number distribution. Higher-order baryon number susceptibilities do increase in proportion to the increasing powers of the correlation length \cite{Stephanov:2008qz}.
Such an increase in correlation length would be reflected in large values of 2nd and higher-order susceptibilities in the vicinity of a critical point and in the region of phase transition. Hence, these quantities are useful indicators of a critical behavior in the CMF model. Deviations of $\chi_n^B$ from the corresponding baselines indicate a transformation between different phases, which is reflected usually in a non-monotonic behavior of these observables, e.g. skewness $\chi_3^B/\chi_2^B$ and kurtosis $\chi_4^B/\chi_2^B$.

The calculated skewness ($\chi_3^B/\chi_2^B$) and kurtosis ($\chi_4^B/\chi_2^B$)~(Fig.~\ref{fig:suscept-diag_muT}) in the CMF model exhibit non-trivial structures in the $T$-$\mu_B$ phase diagram.
The regions of deviations from baseline separate regions with quantitatively different properties which are often dubbed -- ``phases''. Note the sharp phase boundaries indicate first order phase transitions, FOPT, these are only observable at quite moderate temperatures $T<50$ MeV.
The hadron phase located at both low temperatures $T$ and baryon chemical potential $\mu_B$ represents a dilute gas of interacting hadrons.
There, the fluctuation measures  $\chi_2^B/\chi_1^B$, $\chi_3^B/\chi_2^B$, $\chi_4^B/\chi_2^B$ are quite close to unity, consistent with the Skellam distribution baseline. 
The system exhibit a FOPT to a dense hadronic liquid phase with rising chemical potential $
\mu_B\approx1$ GeV. Here, fluctuations are reduced due to the repulsive interactions. Quarks start to appear in moderation. 
The liquid phase exhibits additional FOPT at $\mu_B\approx1.5$ GeV, and second order transition at $\mu_B\approx2.4$ GeV, which show up in the structure of the baryon number susceptibilities at these high $\mu_B$. 
The transition at $\mu_B\approx1.5$ GeV is due to the restoration of chiral symmetry. The transition at $\mu_B\approx2.4$ GeV is due to the quark dominated phase where baryonic density is mainly contributed by quarks. The non-monotonic behavior of the fluctuation measures $\chi_3^B/\chi_2^B$ and $\chi_4^B/\chi_2^B$ reflect the transitions. In contrast to the liquid-vapor transition, those two transitions do not change the Skellam baseline. Hence, the fluctuation measures are rather small $\lesssim 1$ before and after the ``transition'' .

The chiral critical point of the CMF model is located at a rather low temperature $T_{\rm chiral}^{\rm CP}\approx 17$ MeV. This value is close to the critical temperature of the nuclear liquid-gas transition in the same model, the critical $\mu_B^{CP}$ is remarkably different, though: the appearance of the parity partners controls the dynamics of the chiral fields: as the parity partners - in the CMF model - obey the same repulsive interaction strength as the nucleons, the critical point appears at that low temperature. This phenomenon has been observed in various mean field models before.

The different phases shown in Fig.~\ref{fig:order-param_muT} in the $T-\mu_B$ plane are related to the chiral field $\sigma$ and the quark fraction. The chiral field is close to its vacuum value, $\sigma=\sigma_0$, at the hadron gas region, here the quark fraction is close to zero, as expected. Both observables deviate from their vacuum values at higher densities and temperatures only. 

The chiral field drops off more slowly or $\mu_B=0$ than seen in lattice QCD calculations, where chiral field rapidly drops around $T=160$ MeV. The reason for this discrepancy is due to the fact that in the present CMF model the thermodynamics at these temperatures are strongly influenced by baryonic resonances which are not coupled to the chiral fields. Baryon resonances like the $\Delta$ ought to be coupled to chiral fields including their chiral partners. This brings down the chiral condensate of lower temperature as can be seen in \cite{Zschiesche:2004si}.

\begin{figure}[t]
\centering
\includegraphics[width=.5\textwidth]{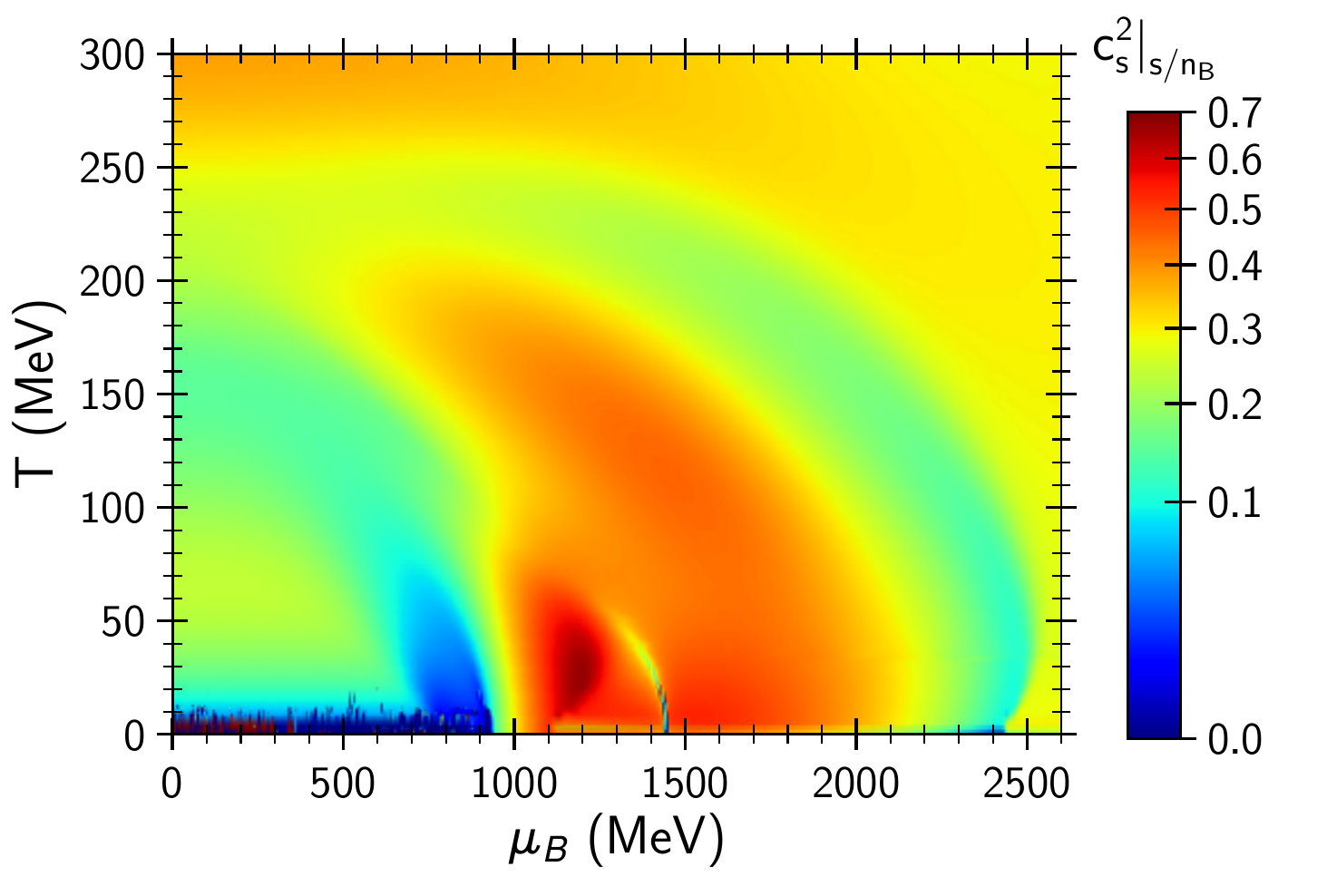}

\caption{The CMF model results for the square of the isentropic speed of sound $c_s^2=\left(\frac{\partial p}{\partial \epsilon}\right)_{s/n_B}$, calculated along lines of constant total entropy density per net baryon density $s/n_B={\rm const}$, as from Eq.~\ref{eq:cs2}.}
\label{fig:cs_muT}
\end{figure}

\begin{figure}[h]
\centering
\includegraphics[width=.5\textwidth]{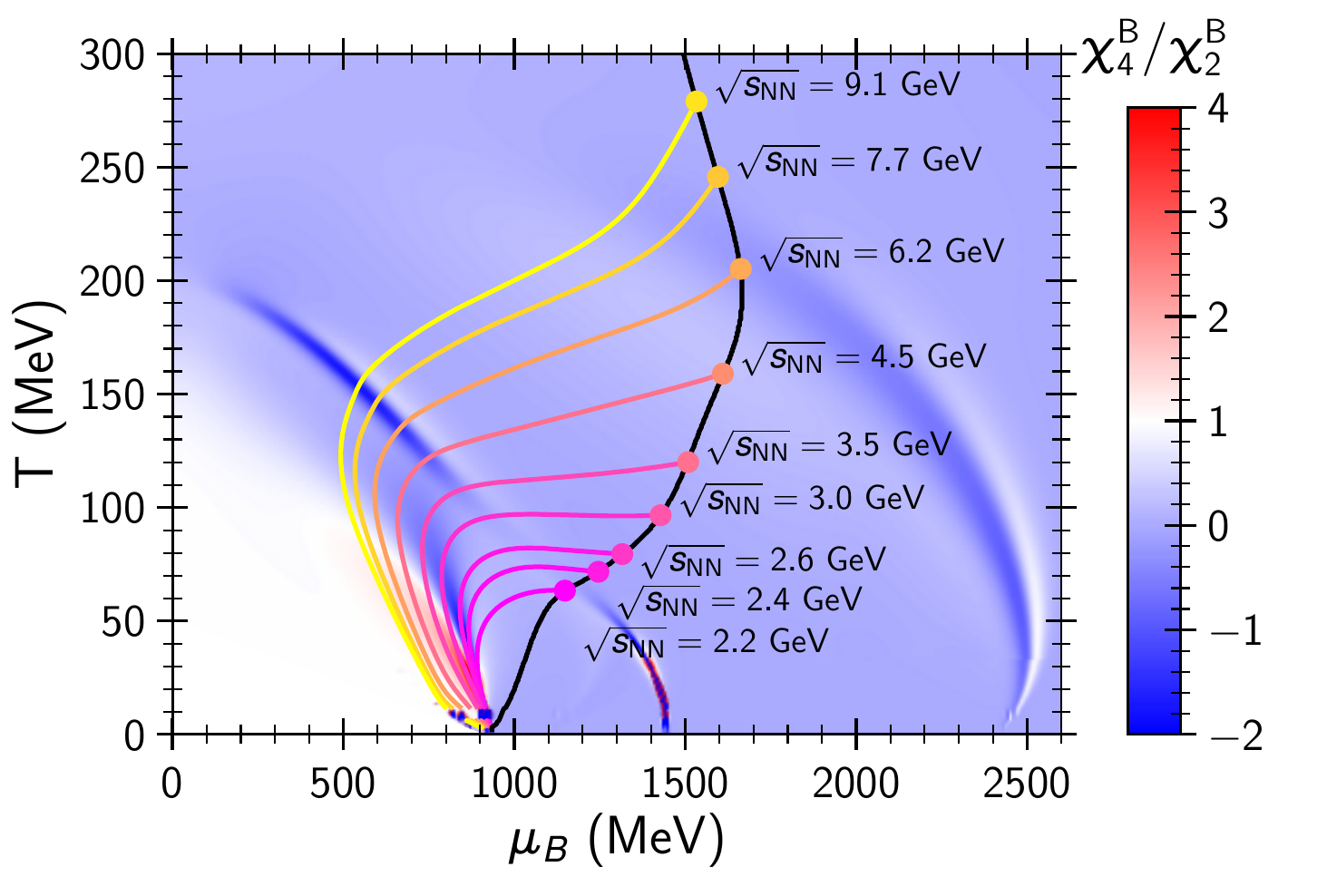}

\caption{Evolution of heavy-ion collisions in the high baryon density region of the  $T-\mu_B$ phase diagram for different collision energies. Black line -- Taub adiabat which describes the initial state of heavy ion collisions as an implicit function of $\sqrt{s_{\rm NN}}$. Colored lines -- isentropic lines of constant entropy per baryon $S/A$ at different bombarding energies $\sqrt{s_{NN}}$ respectively. See Table~\ref{tab:taub} for details.}
\label{fig:isentropes}
\end{figure}

The speed of sound is another important signature for the simulations of the dynamics of heavy ion collisions and neutron star mergers. $c_{s/n_B}^2$ presents a derivative $c_{s/n_B}^2=\frac{\partial P}{\partial \varepsilon}$ at constant $S/A=s/n_B$ entropy per baryon, that allows to estimate a speed of propagation of sound-like excitations in non-dissipative hydrodynamic evolution. The isentropic speed of sound can be calculated as~\cite{1507.05569}:
\begin{equation}
    c_{s/n_B}^2 = \frac{n_B^2\, \partial_T s-2\,n_B\,s\, \partial_{\mu_B} s + s^2\, \partial_{\mu_B} n_B}{(\epsilon + P) \left( \partial_{\mu_B} n_B\, \partial_T s - \left(\partial_T n_B \right)^2\right)}
\label{eq:cs2}    
\end{equation}
The partial derivatives with respect to the chemical potential and to the temperature are performed at constant temperature and at constant chemical potential, respectively. The calculated speed of sound shows three local minima which correspond to three locally softest points of the EoS. These three minima correspond to phase boundaries, where the baryon number susceptibilities present a non-monotonic behavior. Note, that the speed of sound reaches quite large values, $c_s^2\approx0.7$, in the higher density region of nuclear matter. This high speed of sound results due to the strong repulsion between the baryons, before the onset of deconfinement. Thereafter, the vector repulsion and baryon excluded volume cease, as such terms have not been predicted for the quarks.

\begin{table}
    \centering
    \begin{tabular*}{0.45\textwidth}{@{\extracolsep{\fill}} cccccc}
        $\sqrt{s_{NN}}$ (GeV) & $E_{\rm lab}$ (GeV) & $S/A$ & $T$ (MeV) & $n_B/n_0$ &  $\frac13n_q/n_B$\\
        \hline\hline
        2.2 & 1.6 & 2.9 & 63.3 & 2.6 & 0.0 \\
        2.4 & 2.1 & 3.5 & 71.7 & 3.2 & 0.1 \\
        2.6 & 2.7 & 4.0 & 79.5 & 3.7 & 0.2 \\
        3.0 & 3.9 & 5.1 & 96.6 & 4.9 & 0.3 \\
        3.5 & 5.6 & 6.4 & 120.0 & 6.0 & 0.4 \\
        4.5 & 9.9 & 8.3 & 158.9 & 8.6 & 0.7 \\
        6.2 & 19.6 & 10.5 & 205.2 & 14.9 & 0.9 \\
        7.7 & 30.7 & 13.0 & 245.7 & 18.9 & 1.0 \\
        9.1 & 43.2 & 15.0 & 278.9 & 21.3 & 1.0
    \end{tabular*}
    \caption{Initial state properties obtained from the one-dimensional stationary case of the central heavy ion collision -- Taub adiabat~\cite{Taub:1948zz,Thorne:taub_adiabat}. The entropy per baryon $S/A$, temperature $T$, the initial baryon density $n_B/n_0$ and the quark fraction $\frac13n_q/n_B$ are presented for various colliding energies.}
    \label{tab:taub}
\end{table}

\begin{figure*}[t]
\centering
\includegraphics[width=.65\textwidth]{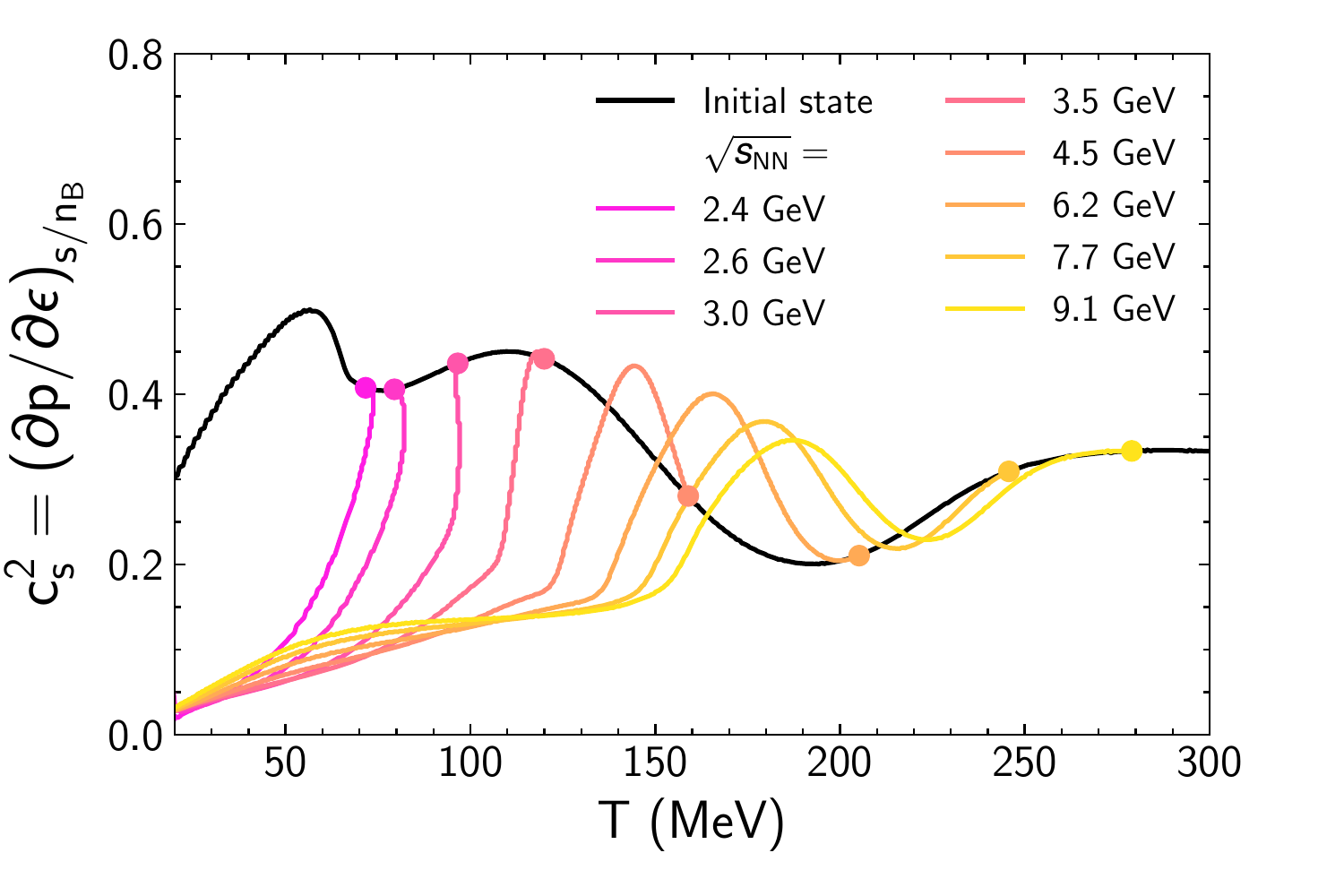}
\includegraphics[width=.65\textwidth]{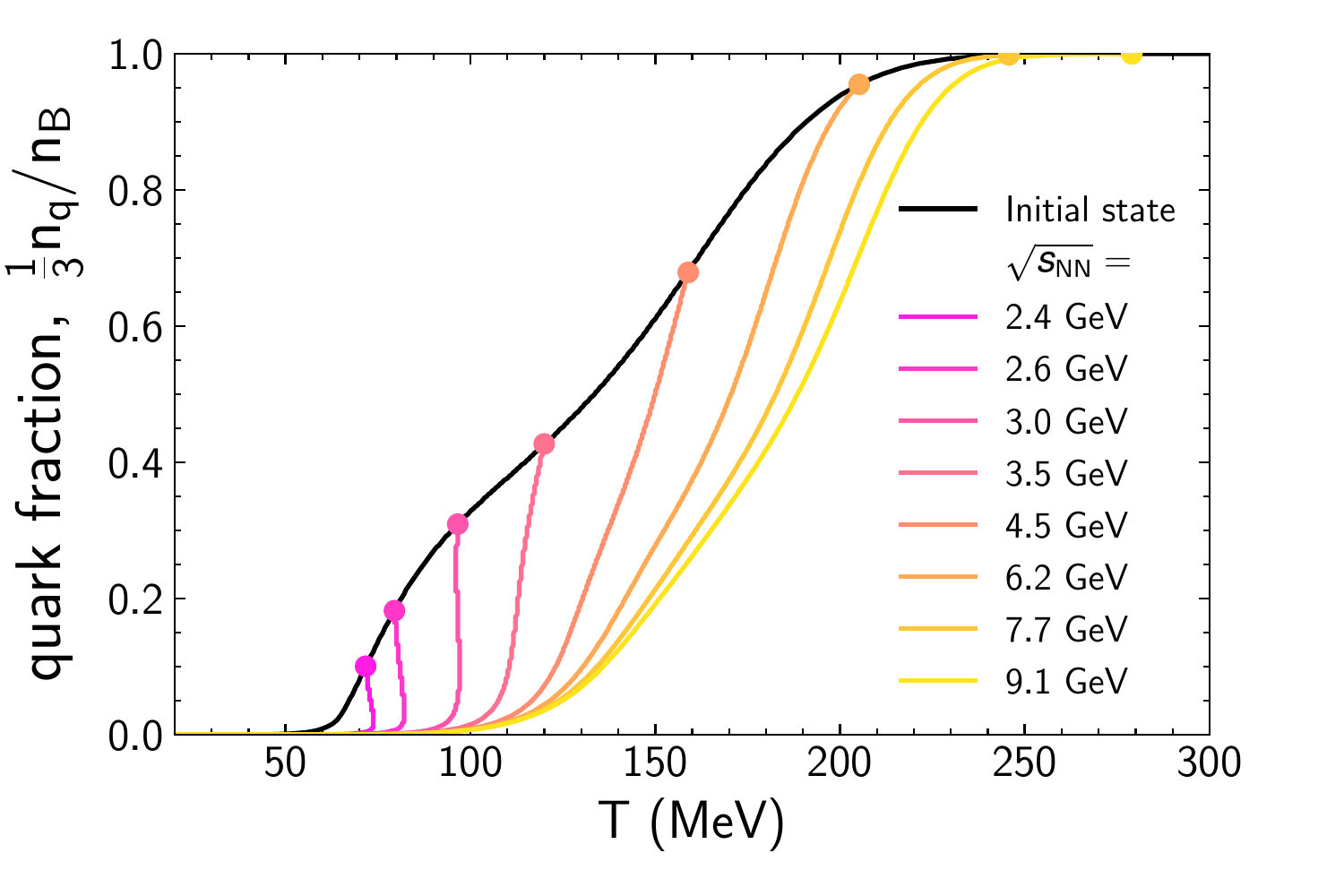}

\caption{Speed of sound at constant entropy per baryon $c_s^2$ ({\bf top}) and quark fraction ({\bf bottom}) along the isentropes as functions of temperature $T$. Colored lines correspond to different collision energies (initial entropy per baryon $S/A$), black solid line correspond to initial speed of sound and quark fraction respectively. Isentropes are the same as in Fig.~\ref{fig:isentropes}. See Table~\ref{tab:taub} for details.}
\label{fig:isentropes_2}
\end{figure*}

\section{Application to heavy-ion collisions}
\label{sec:HI-coll}

The presented EoS is used as input for hydrodynamical simulations of both heavy-ion collisions and neutron star mergers. 
To illustrate which regions of the phase diagram can be reached in collisions at low and moderate collision energies the stationary 1-dimensional Taub adiabat model is used~\cite{Taub:1948zz,Stoecker:1978jr,Thorne:taub_adiabat}. The expansion is described at lines of constant entropy per baryon $S/A=\,$const (isentropes).
These lines depict the isentropic matter evolution of ideal fluid dynamics at different collision energies.

The entropy is produced in the earliest stage of a heavy ion collision by the shock violent compression~\cite{Stoecker:1986ci}. 
During the system's expansion there is only a moderate increase of entropy due to the rather small viscosity~\cite{Csernai:2006zz,Romatschke:2007mq}, hence, an isentropic expansion scenario is a reasonable approximation \cite{Steinheimer:2007iy}.
 
The expansion of the equilibrated matter then continues until the system becomes so dilute that the chemical as well as the kinetic freeze-out occur and the chemical composition is fixed.

The entropy per baryon ($S/A$) is calculated in the 1-dimensional stationary scenario of central heavy ion collisions -- the two colliding slabs of cold nuclear matter \cite{Baumgardt:1975qv,Stoecker:1978jr,Stoecker:1980uk, Stoecker:1981za, Stoecker:1981iw,Hahn:1986mb, Stoecker:1986ci} conserve the baryon number, energy and momentum across the shock front in accord with the relativistic Rankine-Hugoniot equation (Taub adiabat), RRHT,~\cite{Taub:1948zz, Thorne:taub_adiabat}. 
Thus, the produced entropy is directly associated to the collision energy. The thermodynamic properties across the shock front are described by the RRHT-equation
\begin{equation}
    \label{eq:Taub}
    (P_0+\varepsilon_0)\, (P+\varepsilon_0)\, n^2=(P_0+\varepsilon)\, (P+\varepsilon)\, n^2_0\,,
\end{equation}
where $P_0$, $\varepsilon_0$ and $n_0$ correspond to the initial pressure, energy density, and baryon density in the local rest frame of each of the two slabs. The two symmetric slabs consist of the nuclear matter in the ground state, $P_0=0,~\varepsilon_0/n_0 - m_N=-16$ MeV and $n_0=0.16~ {\rm fm^{-3}}$. With any known relation $P=P(\varepsilon, n)$, Eq.~\ref{eq:Taub} can be solved. Furthermore, the collision energy is related to the created density as follows:
\begin{equation}
    \label{eq:stopping}
    \gamma^{\rm CM}=\frac{\varepsilon n_0}{\varepsilon_0 n},~\gamma^{\rm CM}=\sqrt{\frac{1}{2}\left(1+\frac{E_{\rm lab}}{m_N}\right)}\,.
\end{equation}
Here $\gamma^{\rm CM}$ is the Lorentz gamma factor in the center of mass frame of the heavy ion collisions and $E_{\rm lab}$ is the beam energy per nucleon in the laboratory frame of a fixed target collision.
This relation can be obtained from the full stopping condition~\cite{Stoecker:1978jr,Stoecker:1980uk, Stoecker:1981za, Stoecker:1981iw,Hahn:1986mb, Stoecker:1986ci, 1103.3988}. 
The initial state thermodynamics (density, temperature and entropy) of the hot, dense participant matter is obtained from~\cref{eq:Taub,eq:stopping} as a function of the collision energy. The known initial entropy yields  the lines of constant entropy which give the trajectories of the heavy ion collisions in the phase diagram.

The predicted isentropic expansion trajectories are shown in the $T-\mu_B$ plane phase diagram in~\cref{fig:isentropes}. 

Note that 1-dimensional stationary RRHT-adiabat scenario predicts a very strong compression and heating already at intermediate lab (fixed target) bombarding energies. The heavy ion participant system crosses the weak chiral transition predicted by the present CMF model already at $E_{\rm lab}\approx 2$~A\,GeV, i.e. at GSI's SIS18 accelerator facility. Here the specific total entropy is predicted to reach $S/A\approx 3$, in accord with previous RMF-calculations~\cite{Hahn:1986mb} which also used the 1-D RRHT-scenario. The $T-\mu_B$ values, $T\approx 70$ MeV, $\mu_B\approx 1.2$ GeV, with net baryon densities $n_B/n_0\approx 3$, reached here in heavy ion collisions, coincide with the $T-\mu_B$ values reached in binary neutron star collisions,  as recent general relativistic fully 3+1-dimensional hydrodynamical calculations have confirmed~\cite{Hanauske:2017oxo,Most:2018eaw} for the gravitational wave event GW170817. At these temperatures and densities, $T\approx 70$ MeV and $n_B/n_0\approx 3$, the RRHT model predicts that there are abut 20\% of the dense matter are already transformed to quarks.

At $E_{\rm lab}=5.6$~A\,GeV, $\sqrt{s_{\rm NN}}=3.5$~A\,GeV roughly 40\% of the CMF-matter is in the quark state in the RRHT model --  a prerequisite for hot quarkyonic matter. Hence, this energy, which is presumably only reachable by the BMN detector at the Nuclotron at JINR Dubna, and the FXT fixed target setup at the STAR-detector at RHIC, is of great interest: here the matter starts to be dominated by quarks, rather then by in-medium baryons, at $T>100$ MeV and $\mu_B\gtrsim 1.5$ GeV. This is predicted by the present CMF model, when using the 1-D RRHT ideal hydrodynamics. This model predicts that the quarkyonic transition is crossed also at higher energies, using the isentropic expansion of the matter at specific total entropy $S/A>6$. In fact, non-equilibrium viscous effects may increase the specific entropy of the system. However, pre-freezeout radiation, e.g. of Kaons and other hadrons with small scattering cross-sections which can escape early from the semi-equilibrated baryon-rich, dense system can considerably lower the specific entropy during the expansion. So an answer to the question whether the local entropy per baryon increases or decreases during the time evolution awaits more detailed microscopic/macroscopic modeling.

Hence, heavy ions fixed target experiments of SIS at FAiR and SPS at CERN as well as STAR BES program at RHIC probe temperatures from $50<T<280$ MeV and chemical potentials from $500<\mu_B<1700$ MeV for the collision energy range $\sqrt{s_{\rm NN}}<10$ GeV considered here. 
In this region the CMF model shows not an additional phase transition, but the remnants of the nuclear liquid-vapor transition at $T \approx 20$ MeV. The chiral transition at larger chemical potentials can influence the dynamical evolution, too. 
The present results suggest that heavy-ion collisions mostly probe regions where the nuclear matter liquid-vapor critical point dominates -- hence, the observed baryon fluctuations are largely due to remnants of the nuclear liquid-vapor phase transition.  
This had been suggested also in previous works  \cite{Fukushima:2014lfa,Mukherjee:2016nhb,Vovchenko:2016rkn,Vovchenko:2017ayq}.
The CP associated with the chiral symmetry restoration in the CMF model lies at $\mu_B \approx 1.5$ GeV and $T \approx 20$ MeV. 
This high density region is to the best of our knowledge is reachable in the interior of neutron stars, NS, and in binary general relativistic NS mergers \cite{Dietrich:2015iva,Radice:2016rys,Most:2018eaw,Bauswein:2018bma,Hanauske:2019qgs}.

\cref{fig:isentropes_2} presents the square of the isentropic speed of sound, at fixed specific entropy,  $c_s^2$ and the quark fraction as function of the temperature for the studied collision energies, i.e. it shows how respective observable quantities evolve during the cooling of the system while it expands. Those isentropic lines which belong to $\sqrt{s_{NN}}>4.5$ GeV probe the softest point of the EoS, which is attributed to the chiral symmetry restoration. At this energy region there are strong local maxima and minima of the speed of sound squared after which $c_s^2$ rapidly increases during the expansion due to the decrease of the quark fraction, as a result of the rapid appearance of baryons, the EoS stiffens quickly doe to the hard-core baryon-baryon repulsion. For collision energies $\sqrt{s_{NN}}<4.5$ GeV the initial state is not dominated by  quarks, hence, the systems starts to expand at rather high values of $c_s^2$, which then monotonously decrease during the expansion, as a result of the diminishing repulsion between the baryons.

\section{Application to neutron stars}
\label{sec:NS}
 The densities in the neutron star interiors also can exceed by several times the nuclear matter saturation density. At these high densities, the lack of a detailed knowledge of the equation of state and the appropriate microscopic degrees of freedom is similarly disturbing as in the relativistic heavy ion collisions discussed above. The discussion of the role of hyperonic, quarkyonic and strange quark degrees of freedom at these NS densities is ongoing. 

The CMF model can be employed directly to describe the neutron star matter. Here we work out without any changes to the coupling constants and parameters used to describe the $\mu_B=0$ LQCD results. The temperatures isolated in neutron star interiors are negligibly small, in comparison to what we observe in heavy ions collisions and hot QCD scales. The calculations here are done in the limit $T=0$. In contrast to ordinary isospin symmetric nuclear matter, the neutron star matter is in $\beta$-equilibrium which preserves the total electric neutrality of the NS matter and locally ensures stability with respect to $\beta$-decay. As a consequence, strangeness and hypercharges assume finite non-zero values. These constraints require the presence of leptons. In addition, one must allow for quark-hadron degrees of freedom. 

\begin{figure}[t]
  \centering
  \includegraphics[width=0.5\textwidth]{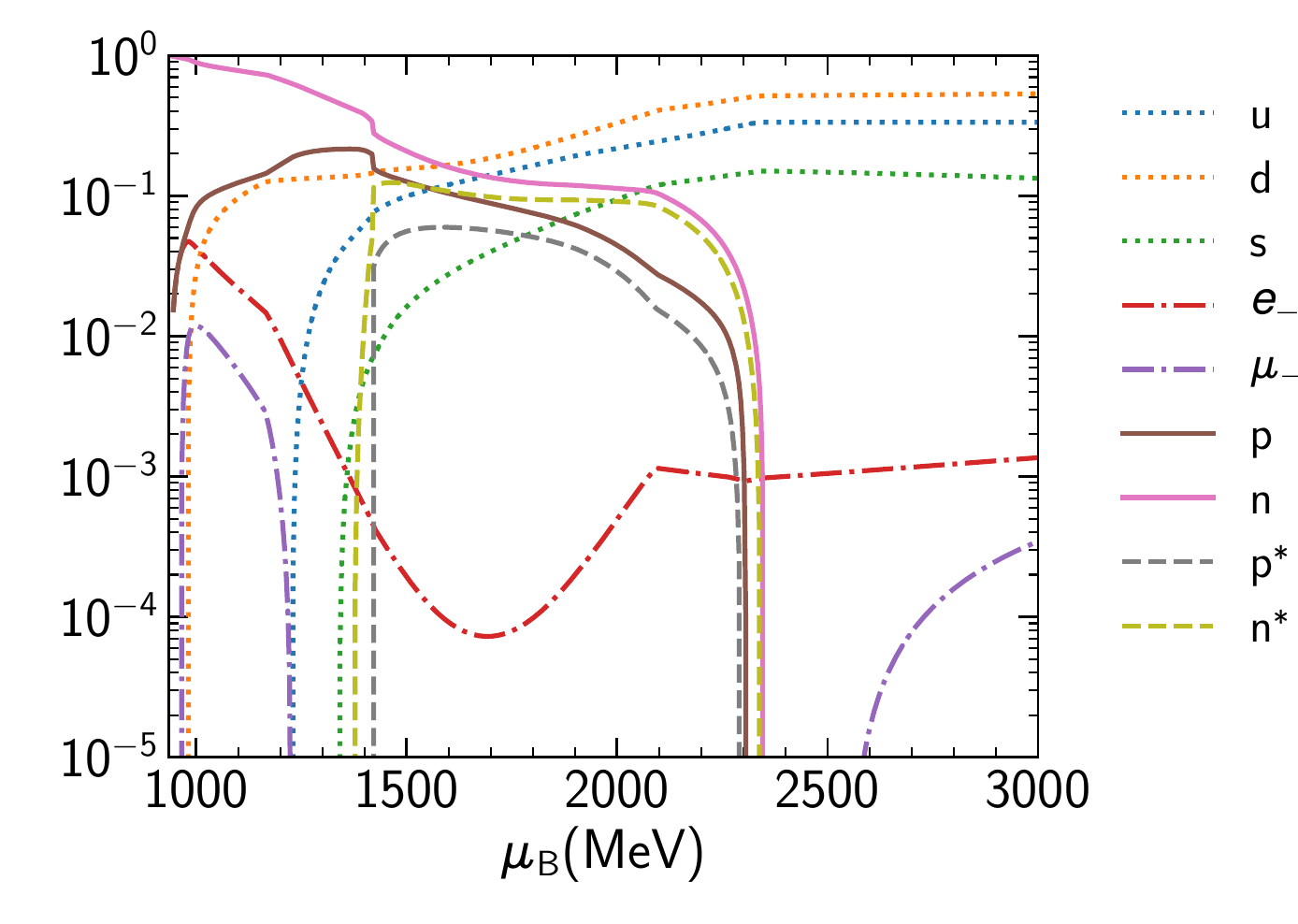}
  \caption{The particle yields, normalized to the baryon density $n_i/n_B$, where a factor of $1/3$ is used for quarks, are shown as the functions of baryon chemical potential $\mu_B$. Dotted lines -- quarks, dash-dotted -- leptons, solid lines -- nucleons, dashed lines -- nucleon parity partners, $p^*$ and $n^*$. The CMF-calculations are performed in $\beta$-equilibrium at $T=0$.
  }
  \label{fig:content}
\end{figure}

Fig.~\ref{fig:content} depicts the CMF model predictions of the relative particle species number content of all different particle species present inside a CMF neutron star, at $T=0$, as function of baryo-chemical potential. One feature of the present CMF model is the absence of baryon resonances (deltas etc. and of hyperons) even though they are included in the CMF model. Their total absence in the present CMF model calculation at $T=0$  is due to  the very strong hard-core repulsion by the excluded volume corrections. 

The calculated EoS at $T=0$ can be used as the input for the Tolman-Oppenheimer-Volkoff (TOV) equation, which allows to relate the mass and the radius of any static, spherical, gravitationally bound object \cite{Tolman:1939jz, Oppenheimer:1939ne}, i.e. here a static neutron star, NS. The outer layer of neutron stars presumably consist of mostly neutron rich nuclei and clusters in chemical and $\beta$- equilibrium. Those nuclei are not yet included in the CMF model. Hence, another input for the EoS of the NS crust is needed. Here, we use the classical crust-EoS~\cite{Baym:1971pw} matched to the CMF-EoS at $n_B\approx0.05~{\rm fm^{-3}}$. 

\begin{figure}[t]
\centering
\includegraphics[width=.5\textwidth]{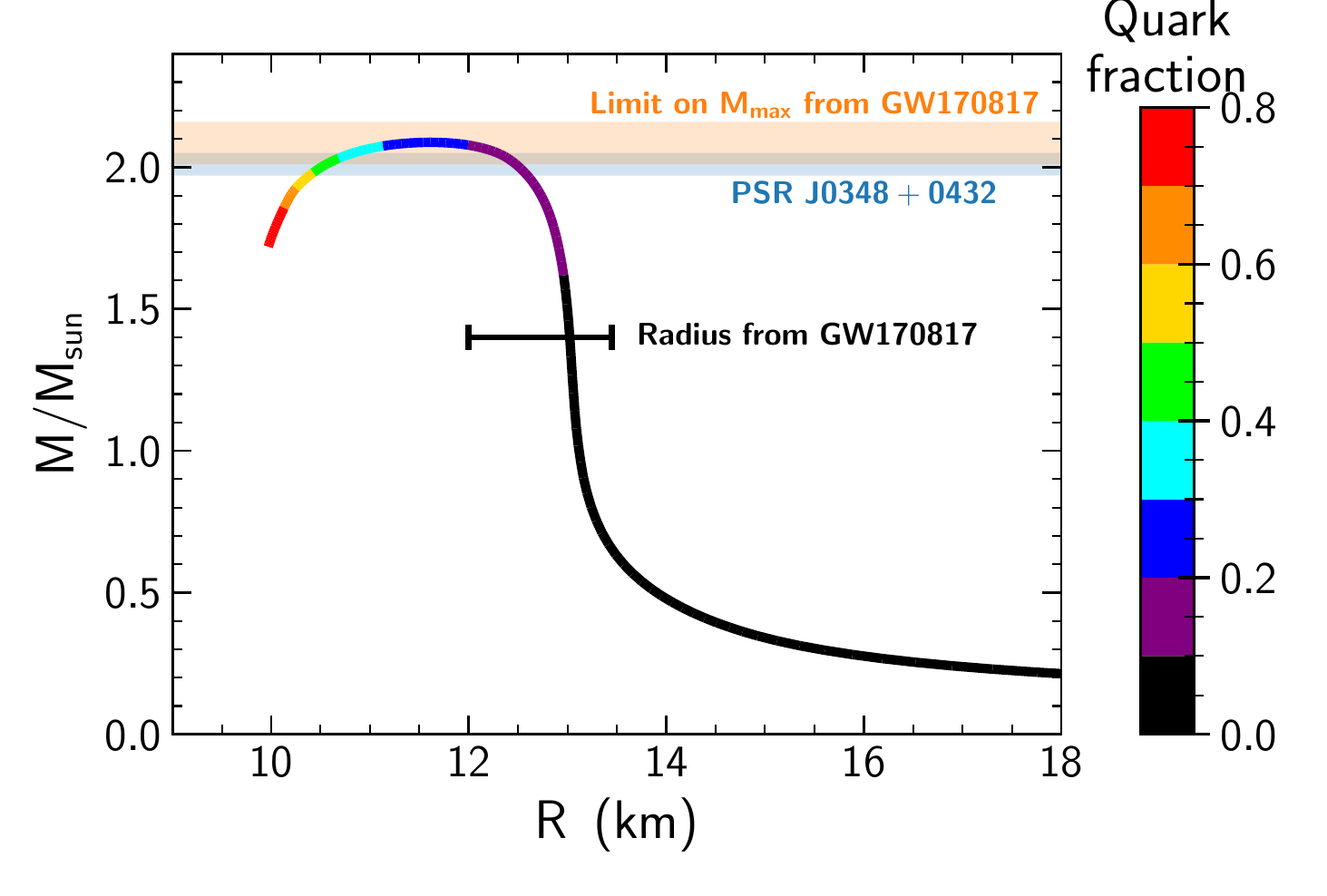}
\includegraphics[width=.5\textwidth]{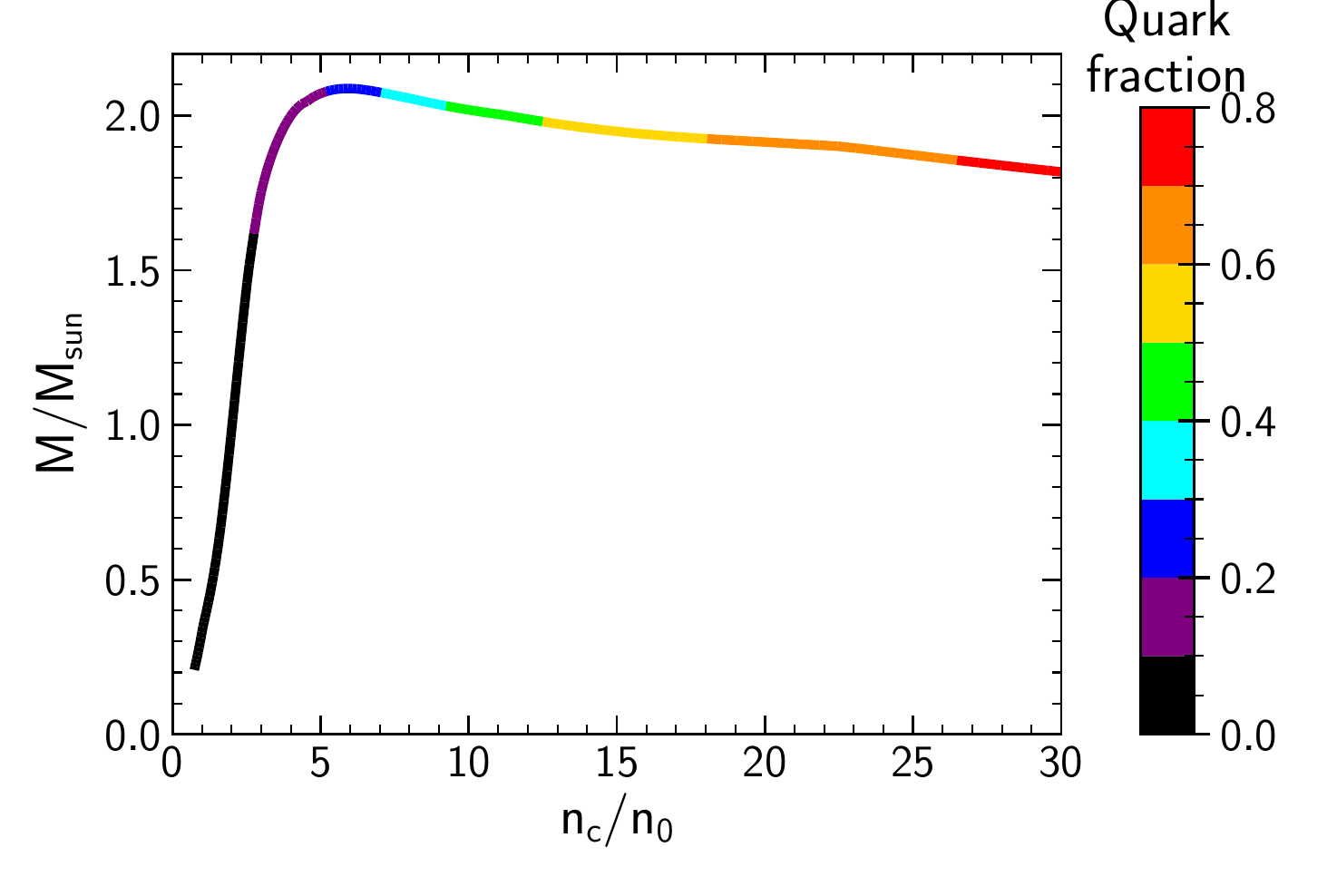}
\caption{The CMF mass-radius diagram ({\bf upper}) for neutron stars is shown as calculated within the CMF model. The CMF neutron star mass is shown as a function of central density ({\bf lower}). Color indicate the fraction of the star coming from the quarks.}
\label{fig:mr}
\end{figure}

\Cref{fig:mr} presents the results on the NS mass-radius relations obtained by solving the TOV equation with the present CMF EoS, matched just to that crust-EoS. The total fraction of the star's mass which consists of light and strange quarks is presented in color-code. The most massive stable solution of the TOV equation contains only $<30\%$ deconfined quarks, i.e. for lighter NS only a small fraction of the star's mass originates from deconfined quark matter. If the quark fraction is increased above $30\%$ the stars become unstable. The central density of the stable stars can never exceed $n_B=6\,n_0$, as shown in the lower part of~\cref{fig:mr}. Here again the maximum mass indicates the ``last stable star''. The continuous slow transition from NS matter to a sizable deconfined quark phase implies a smooth appearance of quarks in the star structure and prevents a ``second family'' of stable solutions to appear. This does prohibit a strict separation between a quark core and the hadronic interior of the star. This is a CMF result due to the Polyakov loop implementation of the deconfinement mechanism. Similar results are obtained in the Quarkyonic Matter-model, where the deconfinement is realized by the appearance of the quarks from inside of the Fermi sea while the hadrons there reside exclusively on the surface shell in momentum space~\cite{McLerran:2018hbz}. A similar approach to deconfinement was suggested in~\cite{Marczenko:2018jui}, however, there the produced mass-radius diagram differs from the CMF-model~\cite{Motornenko:2018hjw} due to the different realization of the chiral symmetry restoration.

The response of a neutron star to non-spherical gravitational fields is reflected in the tidal deformability coefficient $\lambda$~\cite{0911.3535}, which depends strongly on the EoS. During the inspiral phase of binary neutron star merger, both neutron stars experience tidal deformations induced by the other respective accompanying neutron star partner. The tidal deformability $\lambda$ is a measure of the induced quadruple moment $Q_{ij}$ in a response to the external tidal field $\mathcal{E}_{ij}$:
\begin{equation}
    Q_{\ij}=-\lambda \mathcal{E}_{ij}\,.
\end{equation}
$\lambda$ is directly proportional to the second Love number $k_2$:
\begin{equation}
\lambda = \frac{2}{3}k_2 R^5\,.
\end{equation}
For convenience, usually the dimensionless tidal deformability $\Lambda$ is presented as:
\begin{equation}
    \Lambda = \frac{\lambda}{M^5}=\frac{2}{3}k_2\left(\frac{R}{M}\right)^5\,.
\end{equation}
Here, $M$ and $R$ are the mass and radius of the neutron star.
A proper value of $\Lambda$ is important for the description of the inspiral stage during the merger of two neutron stars.

Various estimates of $\Lambda$ emerged after the detection of GW170817 by the LIGO collaboration~\cite{1710.05832}. Ref~\cite{1711.02644} argued that for a $1.4M_\odot$ neutron star the tidal deformability and star radius are constrained to $\Lambda_{1.4M_\odot}>120$ and $R_{1.4M_\odot}<13.6$ km. It was concluded by means of a Bayesian analysis that for a $1.4M_\odot$ star the deformability should be $375.5<\Lambda_{1.4M_\odot}<800$ and the radius is at $12.00 < R_{1.4M_\odot} < 13.45$ km, with respective $2\,\sigma$ confidence levels, see Ref.~\cite{1803.00549}. A recent analysis by the LIGO and Virgo collaborations \cite{1805.11579,1805.11581} provides detailed constraints, by using a Bayesian analysis based on the reproducing of the details of the gravitational wave signal.

\begin{figure}[t]
\centering
\includegraphics[width=.5\textwidth]{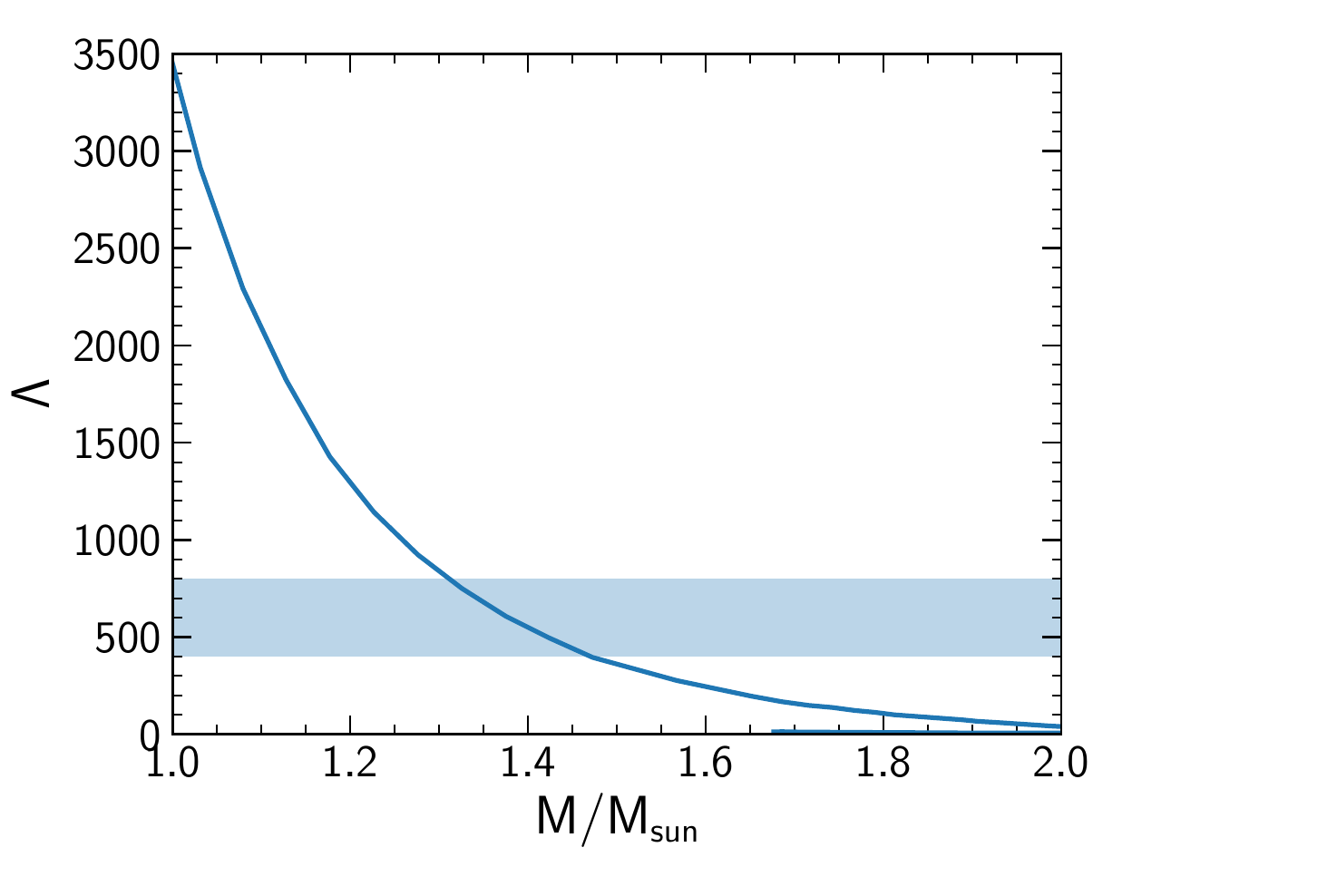}
\includegraphics[width=.5\textwidth]{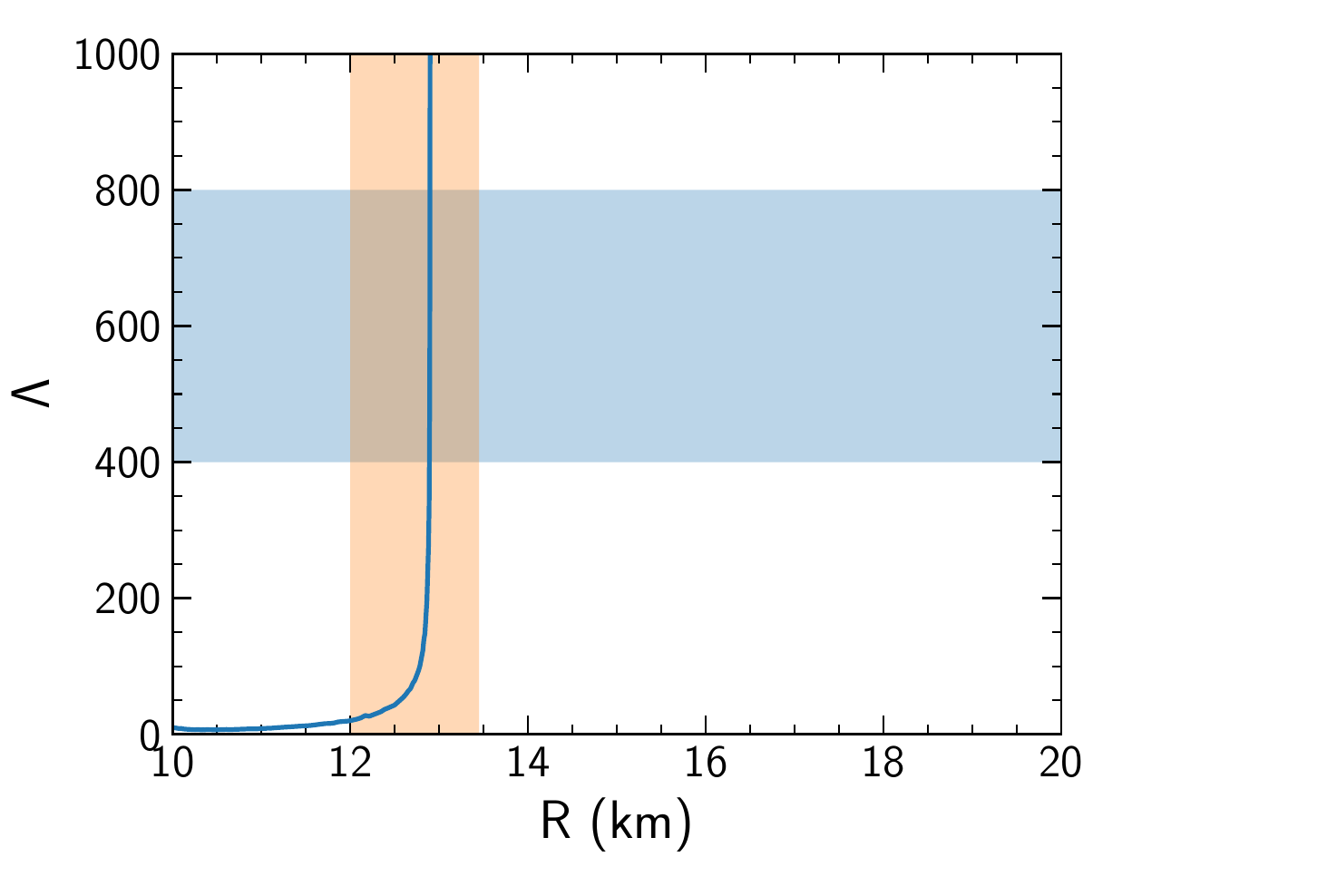}
\caption{Tidal deformability $\Lambda$ as function of NS mass ({\bf upper}) and radii ({\bf lower}). Blue bands correspond to $\Lambda$ constraints of NS with $M=1.4~M_{\rm sun}$, and yellow bands -- constraints on radius of NS with $M=1.4~M_{\rm sun}$ \cite{Most:2018eaw}.}
\end{figure}

\section{Summary}
\label{sec:summary}
A unified and consistent approach to QCD thermodynamics is presented which is appropriate for essentially all temperatures and densities relevant for both heavy ion collisions and for neutron star matter. The Chiral SU(3)-flavor parity-doublet Polyakov-loop quark-hadron mean-field model, CMF model, includes the main features of QCD hadron phenomenology, as well as a very good description of known QCD thermodynamics. The CMF model allows for a simultaneous description of many nuclear (astro-) physical data, consistent with astrophysical observations as well as heavy ion collisions and of compact stars. The CMF model is improved by the fixing of the relevant CMF parameters for the quark sector to the state-of-the-art LQCD data on the interaction measure. Here, the parameters of the Polyakov-loop potential and of the quark couplings to the chiral fields have been fixed. A good agreement is found between the CMF model predictions for LQCD data on both the baryon number susceptibilities and the ``lines of constant physics''. The CMF model is used to explore the phase diagram of strongly-interacting matter at a wide range of $T$ and $\mu_B$. Three critical regions are found, which are connected to the nuclear liquid-vapor phase transition, to the chiral symmetry restoration, and to the deconfinement. The region of phase diagram accessible to experiments of high energy heavy-ions collisions is dominated by remnants of the nuclear liquid-vapor phase transition. Other critical regions may be probed by neutron star structure and in binary neutron star mergers. The calculated properties of neutron stars, like the mass-radius relation, the chemical composition of the stars and the tidal deformabilities are  in a good agreement with recent experimental observations. The applicability of the improved CMF model to such a wide range of strongly interacting systems is impressive. For the first time, a QCD-motivated EoS is presented which precisely describes the thermodynamic observables for the whole QCD phase diagram.

\section*{Acknowledgements}
\label{sec:acknowledgements}
The authors thank HIC for FAIR, HGS-HIRe for FAIR, BMBF, and DFG for support.
J.S. appreciates the support of the SAMSON AG, WGG-Forderverein, and the C.W. F\"uck-Stiftungs Prize 2018. H.St. acknowledges the support through the Judah M. Eisenberg Laureatus Chair at Goethe University, and the Walter Greiner Gesellschaft, Frankfurt. Computational resources have been provided by the Center for Scientific Computing (CSC) at the J. W. Goethe-University, Frankfurt. 

\bibliography{QCD_thermodynamics}

\end{document}